\documentclass[aps,pre,showpacs,showkeys,amssymb,amsmath,superscriptaddress,reprint]{revtex4-1}

\usepackage[pdfpagemode=None,pdfstartview=FitH]{hyperref}
\usepackage{amsthm}
\usepackage{amsmath}
\usepackage{graphicx}
\usepackage{color}
\usepackage{verbatim}
\usepackage{subfigure}
\usepackage{amssymb}
\usepackage{bm}
\usepackage{hhline}
\usepackage{gensymb}
\begin{document}

%\title{Capture dynamics of inert colloids in the presence of chemically active colloid}
\title{Formation of self-propelling clusters starting from randomly dispersed Brownian particles}

%Formation of spontaneous chemical motors by diffusiophoretic capture of inert colloids to a chemically active colloid}
\author{Prabha Chuphal}
\affiliation{Department of Physics, Indian Institute of Science Education and Research, Bhopal, India}

\author{Ishwar Venugopal}
\affiliation{Department of Physics, Indian Institute of Science Education and Research, Bhopal, India}

\author{Snigdha Thakur}
\affiliation{Department of Physics, Indian Institute of Science Education and Research, Bhopal, India}

%\date{\today}

\begin{abstract}

We present a simple chemical strategy for the formation of a self-propelling cluster via the process of capture and assembly of passive colloids on the surface of a chemically active colloid. The two species of colloids that are isotropic and Brownian otherwise interact to form propelling cluster.  With the help of coarse-grained numerical simulations, we show that a chemically active colloid can induce diffusiophoretic motility to nearby chemically inert colloids towards itself. This propulsion and then self-assembly can then lead to the formation of active cluster. We observe the formation of propelling dimers, trimers, tetramers, etc. depending on the chemical activity and size of the colloids. 
\end{abstract}

\maketitle

\section{Introduction}
\label{sec:introduction}

Both biological and non-biological organisms have been reported to respond to the chemical gradients around them, by moving either up or down the gradient. In biological framework, the bacteria have been found to migrate towards the glucose-rich regions in search of their food and form active assemblies in this process~\cite{Zhang:2010}. Sperm cells are also well known to collectively swim towards the chemoattractants~\cite{Eisenbach:2006}. On the other hand, in the realm of synthetic nano/micro world, the colloidal particles have been designed to act like artificial microswimmers~\cite{Li:2005,Gao:2013,Sundararajan:2008,Guix:2014,Yu:2018}. These microswimmers have enormous potential to understand many elementary questions in active matter and self-organization~\cite{SY:2018,Yu:2018,Palacci:2013,Schmidt:2019,Gong:2019}. Also can be employed as carriers to pick-up and deliver cargo~\cite{Oerlemans:2010,Gao:2013,Sundararajan:2008,Yu:2018,Alapan:2018}.

In the view of having a variety of applications of such assembled structures in real and synthetic life, they have attracted remarkable attention in past decade~\cite{Zhang:2010,Yan:2012,Palacci:2013,SY:2018,Yu:2018,Gong:2019,Schmidt:2019}. These studies have led to the creation of various functional materials as per the demand of present technologies. Dynamic self-assembly in a mixture of eccentric active particles and the passive particles was shown to form a large dense, dynamic cluster~\cite{Ma:2017}. Similarly, a colloidal suspension of magnetic particles was demonstrated to exhibit dynamic assemblies depending on the applied magnetic field frequencies~\cite{Martin:2013}. Recent experimental work on a suspension of light-absorbing and non-absorbing immotile microspheres have shown to form an assembly of active colloidal molecules in a highly controlled manner using light~\cite{Schmidt:2019}. In the context of chemically active colloidal particles self-assembly of a pair of colloid leading to the spontaneous motor formation was also revealed in recent studies~\cite{SY:2018,Yu:2018}. Similarly, chemically modifies Janus has been illustrated to form on-the-fly assemblies that have potential application toward optimal cargo transport and delivery~\cite{Gao:2013,Takahara:2005,Bogart:2014}. During the process of self-assembly, the role played by an individual entity is important. In this work, our interest lies in the process where nonequilibrium interaction between the colloids leads to assemblies that can spontaneously acquire directional motility.  The significant challenges in this regard are to fine-tune the factors that influence the assembly and motion. Some of the factors which ultimately decide the structure and then the motion of assembled structure are the motility of participating entities~\cite{Whitesides:2002} which can be influenced by the generated chemical gradient~\cite{Moran:2017,Yu:2018}, thermal gradient~\cite{Kroy:2016}, externally applied fields~\cite{Yan:2012,Martin:2013}, etc. Also, the hydrodynamic interactions present in the system are crucial to determine the same~\cite{Marchetti:2013,Singh:2016,Shen:2019}.

Here, we study the dynamics of self-assembly followed by self-propulsion in a system having two species of colloids. Both of these isotropic colloids are diffusive individually, but upon interaction, they acquire motility. Out of these two species, one is chemically active colloid that form a uniform radial chemical gradient when immersed in a chemical fuel. The other class of colloid that we name as an inert or passive colloid senses chemical gradient produced by the active sphere and exhibit self-propulsion towards it, like a living bacteria move towards chemo-attractants. Such combination of chemically active and passive spheres have been studied in the past in the context of propelling dimers~\cite{Gunnar:2007,Valadares:2010,SY:2018}. With the help of mesoscopic coarse-grained simulation methods, we show that inert colloids in the neighbourhood of a chemically active colloid get attracted towards it through diffiosiophoresis. Hence get captured on the surface of active colloid thus making an assembly. Another interesting observation is the reconfiguration of the assembly, \textit{i.e.} the spontaneous rearrangement of inert spheres on the surface of active particle such that the cluster then starts to swim.  The active assembly of the colloids in the form of a dimer, trimer, tetramer, etc. is observed in our study depending on the radius of active and inert colloids. The propulsion speed, capture time, and assembly time have been quantified here, and the factors affecting the performance of the motors have been identified. 

The article is organized as follows. In section ~\ref{model}, we have described the simulation model and the system of active and passive colloids in the chemically active medium. Section~\ref{results} discusses the capture dynamics of inert particles on the surface of active colloid and also explains the reason for self-assembly and propulsion. In this section, we start the discussion with a pair of colloids and then move to multiple inert particle system. Finally, the section~\ref{conclusion} concludes the study.

\section{Diffusiopheretic theory and Simulation model}\label{model}

\subsection{Simulation}
We employ a particle-based hybrid simulation model where the particle motion is governed by the molecular dynamics (MD) coupled with Multi-particle collision dynamics (MPCD) to implement the hydrodynamics and thermal effects. This is a well-known method and has been used previously in different contexts~\cite{Gunnar:2007,Valadares:2010,SY:2018}.  Initially a reactive colloid $(R)$ with radius $R_{r}$ and inert colloids $(I)$ with radius $R_i$ are immersed in a solvent with $A$ type point-particles. The reactive colloid $(R)$ is a chemically active species and converts the fuel fluid particle $A$ to the product particle $B$ by an irreversible chemical reaction $A + R \rightarrow B + R$ with an intrinsic rate. On the other hand, the chemically inert colloids $(I)$ do not possess this property. The chemical reaction on $R$ colloid hence generates a radially outward gradient of product molecules, which can be sensed by any inert colloid $I$ in its vicinity. These inert colloids $I$ hence exhibit the diffusiophoretic motion~\cite{Anderson:1989} towards $R$ upon experiencing the chemical gradient.

\begin{figure} [h!]
	\centering
	\includegraphics[width=0.7\linewidth]{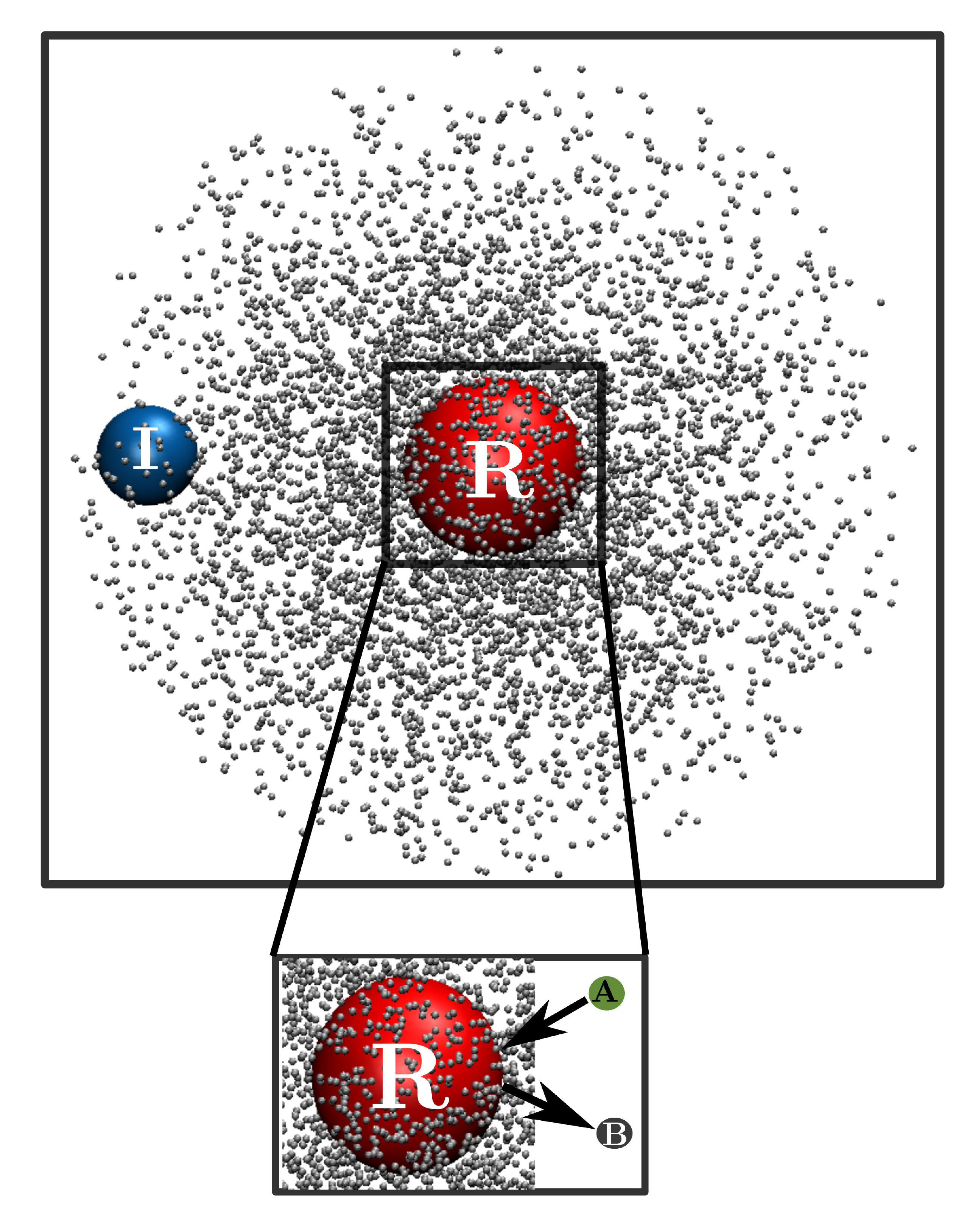}
	\caption{Schematic for the conversion of A-type solvent particles to B-type at the surface of the reactive $(R)$ sphere. Inert $(I)$ sphere does not take part in any chemical reaction.}
	\label{fig:sch}
\end{figure}

 The colloidal spheres interact with the fluid particles through the repulsive Lennard-Jones(LJ) potential given as,
$V_{lj}(r) = 4\epsilon
[(\frac{\sigma}{r})^{12}-(\frac{\sigma}{r})^{6}] + \epsilon$  for $r \le r_{c}$ and $V_{lj}(r) = 0$ for $r > r_c$. Here $\sigma$ and $\epsilon$ are the characteristic distance and energy parameter. To model the diffusiophoretic motion to the colloids, we choose the interaction energy parameter of $A$ and $B$ as $\epsilon_A = \epsilon_B = \epsilon$ for $R$ sphere. Whereas for $I$ spheres we choose $\epsilon_B < \epsilon_A = \epsilon$. In addition to this, excluded volume interaction has been employed between two colloids. The combination of chemical gradient produced by $R$ sphere around $I$ (see Fig.~\ref{fig:sch}) along with the above stated difference in interaction energies of fuel and product with $I$ leads to the propulsion of $I$ spheres towards the $R$ sphere. To keep the system out-of-equilibrium, which is required for continuous propulsion of $I$, the product $B$ particles are converted back to $A$ at a distance far from spheres.

We use a coarse-grained mesoscopic dynamical model for fluid particles, popularly known as multi-particle collision dynamics(MPCD)~\cite{Raymond:2008,Gompper:2009}. In this scheme, $N_{s}$ point-like particles with mass $m_{s}$ can move in a cubic box of length $L$. The method is comprised of two alternative steps; streaming and collision. With the help of forces based on the potentials used in the system, the fluid particles are evolved by Newton's equation of motion in the streaming step. While, in the collision step, all the fluid particles are sorted into small cubic cells of size $a_{0}$ such that the mean free path $\lambda<a_{0}$. The multi-particle collision is performed in each cell by a random rotation matrix $\mathcal{R}$ that results in the rotation of the relative velocities of all the fluid particles. The velocity of fluid-particle $i$ after each collision step $\tau_{c}$ in cell $\chi$ is given by
${\mathbf{v}}_{i}(t+\tau_{c})={\mathbf{V}}_{\chi}(t)+{\mathcal{R}}  [\mathbf{v}_{i}(t)- {\mathbf{V}}_{\chi}(t)]$

where $\mathbf{V}_{\chi}$ is the velocity of the center of mass of all fluid particles in cell $\chi$. A random grid shift in each direction is applied to ensure the Galilean invariance for the system ~\cite{Ihle:2001}. The described method here conserves mass, momentum, and energy.

{\bf Simulation Parameters:} Throughout the paper, all the parameters are described in dimensionless units $a_{0}$, $\epsilon$ and $m_{s}$ for length, energy, and mass respectively. We use the cubic simulation box of dimension $L=50$. The temperature of the system is fixed at $\kappa_{B}T=1.0$. Unit cell length $a_{0}$ is taken for performing the multi-particle collisions. In each MPC cell, the velocity rotation is carried out by an angle of $\alpha=120{\degree}$ about a random axis at every collision time step $\tau_{c}=0.1$. Average fluid number density in each MPC cell is $c_{0}=10$, and the mass of fluid-particle is taken to be $m_{s}=1.0$. Energy parameters in LJ potential  $\epsilon=1.0$ whereas $\epsilon_{B}$ is varied as $0.01, 0.1$, and $0.5$ to achieve different propulsion rates of $I$ colloids. To study the size effect, we have taken two different sizes of the reactive colloid  $R_{r}=2.0$ and $R_{r}=4.0$. Whereas, the size of the inert colloid is kept fix at $R_{i}=2.0$. The masses of colloidal particles are adjusted according to the density matching with the surrounding fluid. MD time step is taken to be $\Delta{t}=0.01$.

\subsection{Diffusiophoretic theory for the inert colloid}

The inert colloid $I$ when comes in the vicinity of $R$ colloid, can feel the gradient of $B$ type particles around its surface and respond to it by moving towards the $R$ colloid following the diffusiophoresis mechanism~\cite{Anderson:1989}. The different interaction of $A$ and $B$ type particles with the surface of inert colloid makes it move due to the fluid slip velocity $v_s$ generated around the boundary layer~\cite{Golestanian:2007}:

\begin{equation}
v_{s}=-\frac{k_{\beta}T}{\eta}(\nabla C_{B})\Lambda,
\end{equation}

where $\eta$ is the viscosity of the medium, $C_{B}$ is the
concentration of product $B$ particles on the outer edge of the boundary
layer, and $\Lambda$ determines the net strength of interaction between the
 fluid particles and the surface of the colloid. The velocity of
inert colloid can be calculated by averaging the slip velocity over the
entire surface of the colloidal particle~\cite{SY:2018}. To do this an estimation the concentration field $C_{B}$ is required, which can be evaluated by solving the diffusion equation with the help of radiation boundary condition~\cite{SY:2018}. The final directed velocity of $I$ can then be given as:

\begin{equation}
V(t) =  \frac{2 k_{B} T C_0 \Lambda}{3 \eta} \frac{k_{0}}{k_{0} + k_{D}} \frac{R_r}{R(t)^2} \equiv \frac{\lambda}{R(t)^2}, 
\label{eq:vz}
\end{equation}

where $\Lambda =\int_{0}^{\infty} dr \: r (e^{-\beta V_{B}(r)} - e^{-\beta V_{A}(r)} )$, $k_{D}=4\pi R_{r}D$ is Smoluchowski rate constant, $\kappa_{0}$ is intrinsic reaction rate constant and $D$ is the diffusion coefficient of the fluid particles. $R(t)$ is the distance between $I$ and $R$ at a given time.

Further, the time evolution of separation between the colloidal particles can be obtained by integrating Eq.~\ref{eq:vz} and the time taken by inert colloid to reach the encounter distance $R_f = R_{r}+R_{i}$ is given by,

\begin{equation}
\langle \tau_{cap} \rangle=\frac{(L_{d}^{3}-R_{f}^{3})}{3 \lambda}.
\label{eq:taucap}
\end{equation}

\section{Results and Discussions}\label{results}

The interaction between a pair of freely moving $R$ and $I$ colloid having different sizes has been investigated earlier both experimentally and theoretically~\cite{Yu:2018,SY:2018,Reigh:2015}. Here, our interest is to explore what happens to the dynamics of $R$ colloid in the presence of multiple $I$ spheres. To be precise, we want to probe the capture and assembly dynamics of $I$ sphere on $R$, which then leads to symmetry breaking and hence converts a diffusive sphere to a chemical motor. To begin with, let us re-examine the results from the case of $R$-$I$ pair.

\subsection{Capture dynamics of a single inert colloid}\label{subsec1}

\begin{figure} [!h]
	\centering
	\includegraphics[width=1.0\linewidth]{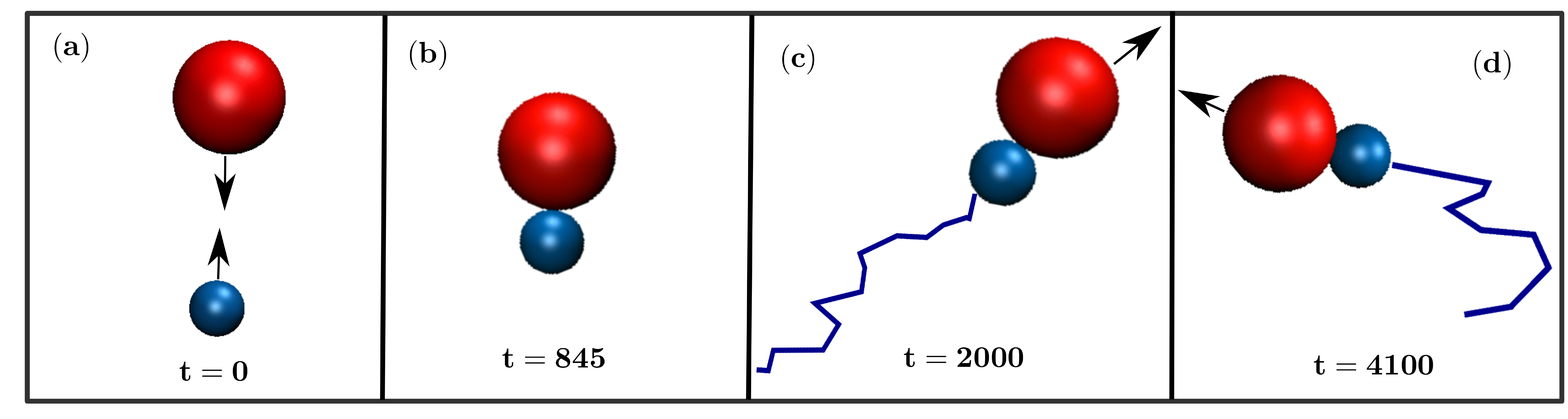}
	\caption{Dimer motion with representative trajectories when $R_{r}=4.0$ and $R_{i}=2.0$.}
	\label{fig:color1}
\end{figure}

When the chemical reaction on $R$ sphere is switched on, a chemical gradient is produced around $I$, which then starts moving towards $R$ by diffusiophoresis and eventually gets captured. In the entire process of capture, $R$ sphere mostly exhibit diffusive motion, which then transforms to weak drift towards $I$ when they are very close to each other.  A stable dimer is formed once the $I$ and $R$ sphere come very close to each other, and the assembly then exhibits a directional motion over a long distance and time similar to a motor. The motion of this bound pair is similar to the sphere-dimer motors~\cite{Yu:2018}. Fig.~\ref{fig:color1} shows the snapshots from simulation for capture dynamics and directional motion of dimer.

\begin{figure} [h!]
	\centering
	\includegraphics[width=1.0\linewidth]{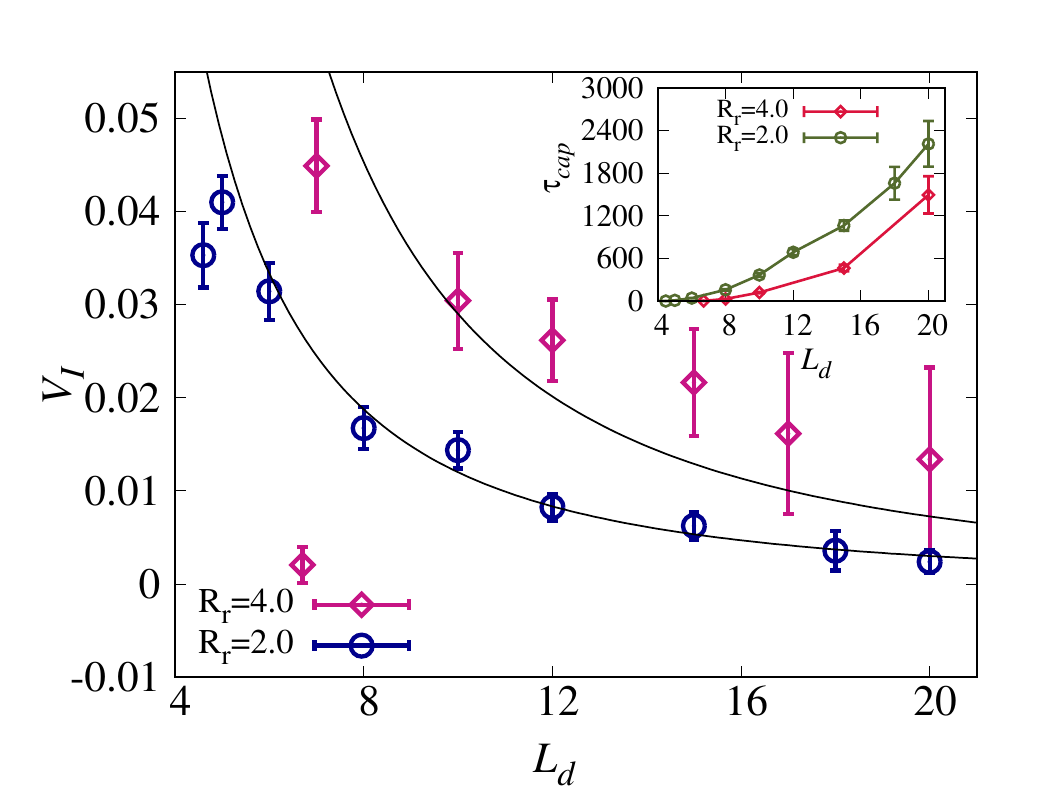}
	\caption{Propulsion velocity $(V_{I})$ of inert colloid varying with initial separation $L_{d}$ for two sets of radii of reactive colloids, $R_{r}=4.0$(purple plot) and $R_{r}=2.0$(blue plot) and average capture time ($\tau_{cap}$) taken by inert colloid is shown in inset.}
	\label{fig:vz_tau}
\end{figure}

Chemotaxis and capture of inert colloid to the reactive target site is quite robust and seems to take place for various sizes of colloids. Of course when the size of $I$ becomes too small, thermal fluctuations disrupt the directed motion of $I$ sphere. On the other hand, the radius of $R$ sphere plays an important role in both capture and propulsion.  The concentration field of $B$ species around $I$ which is a key factor in determining the velocity and capture time, is influenced by $R_r$. Therefore, we probe the effect of size of $R$ on the capture dynamics. The velocity of $I$ sphere $V_I$ is plotted in Fig.~\ref{fig:vz_tau} as a function of  $L_d$, which is the separation distance between the center of spheres with $R_{r}=2.0$ and $R_{r}=4.0$ respectively. As the distance between the spheres decreases, the concentration gradient of $B$ around $I$ increases leading to increased velocity. Furthermore, the propulsion velocity $V_I$ is substantially more for larger $R_r$, which is in line with Eq.~\ref{eq:vz}. The capture time also then reduces accordingly. For all later studies, since the focus is on moving $I$, therefore we choose $R=4.0$.

\subsection{Capture dynamics for a pair of inert colloid}\label{sub1}

\begin{figure} [h!]
 	\centering
 	\includegraphics[width=0.4\linewidth]{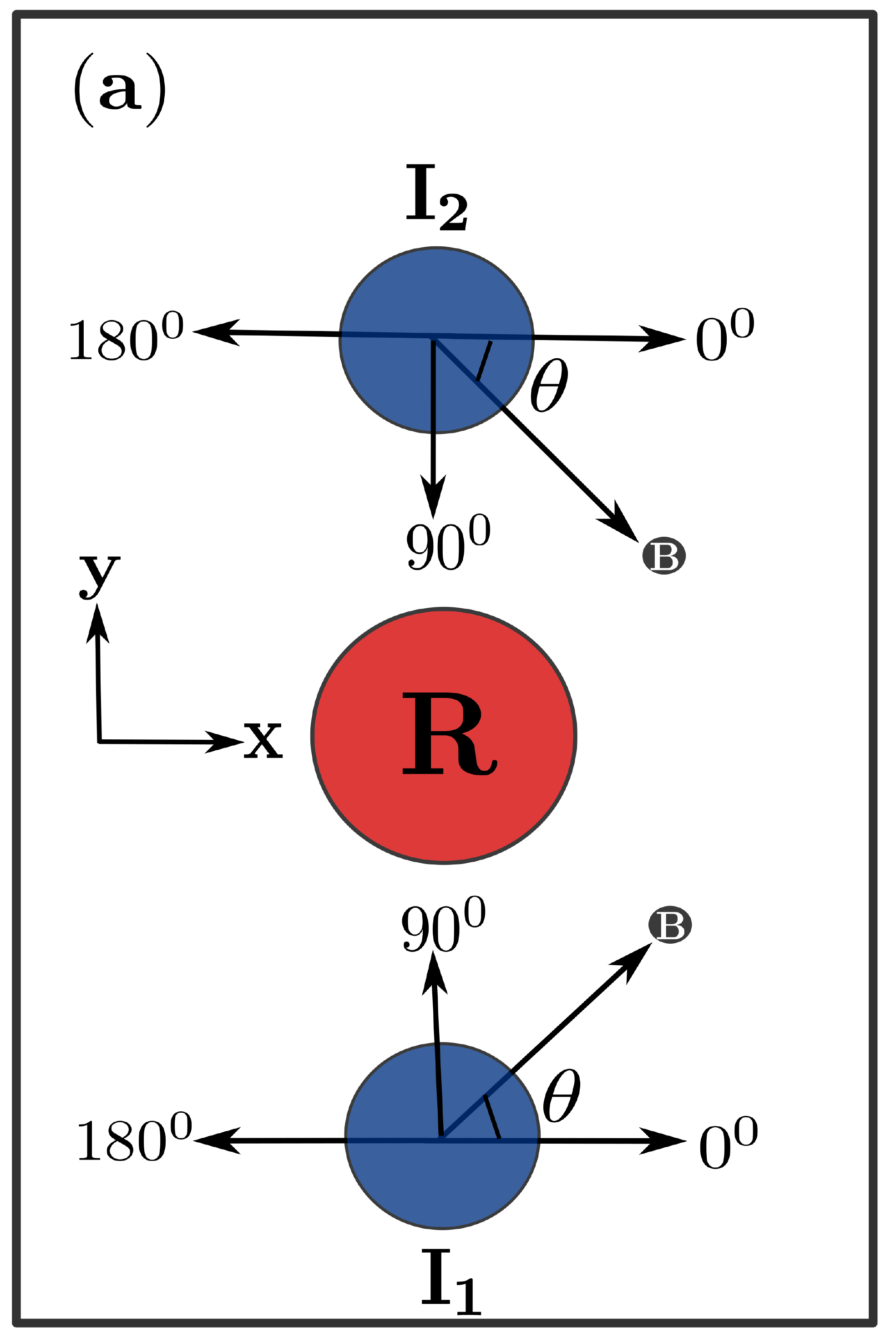}
 	\includegraphics[width=0.5\linewidth]{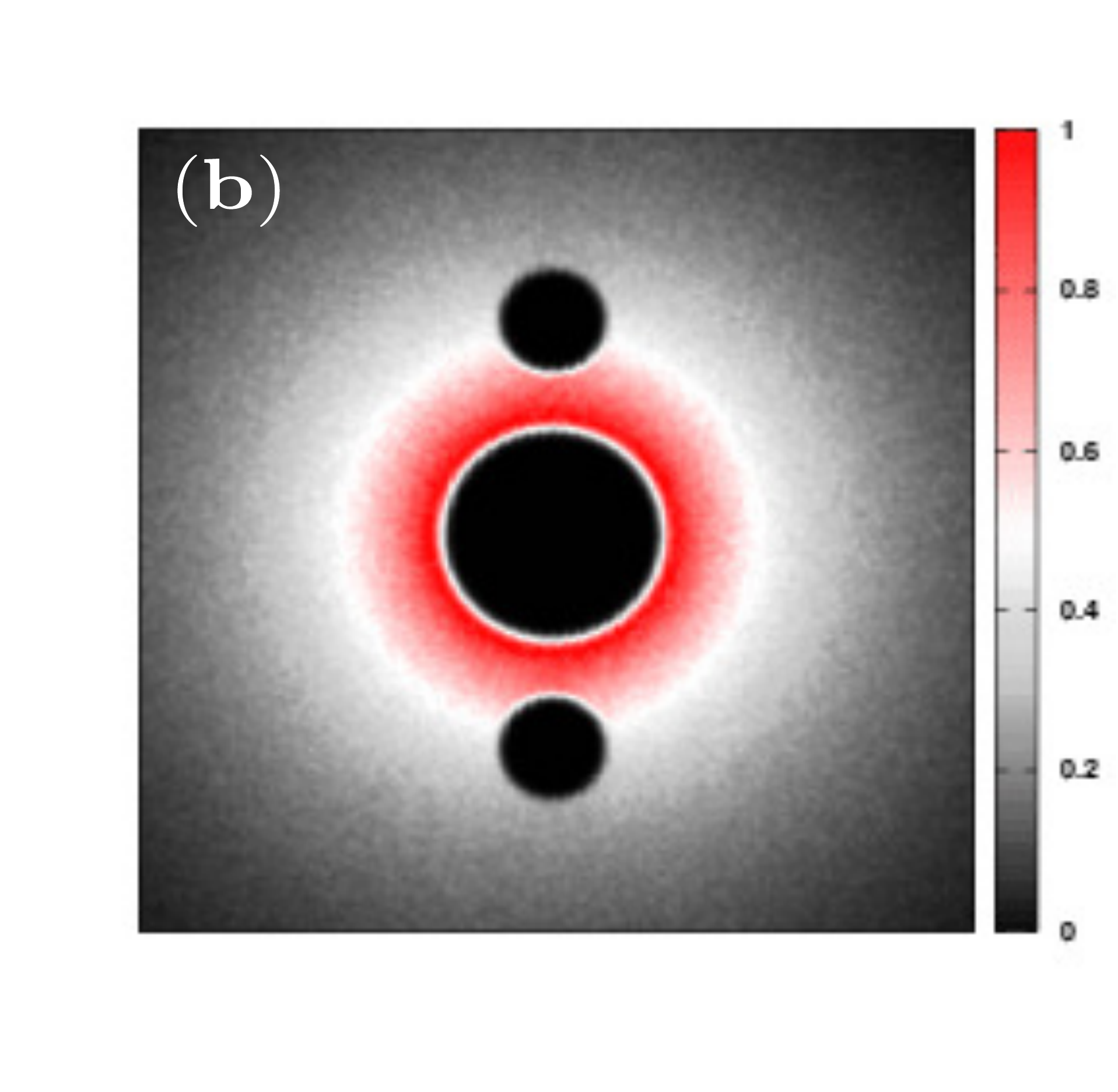}
 	\caption{ (a) A schematic representation of the initial configuration of the system with two inert spheres $(I_{1}$ and $I{2})$ and a reactive sphere $R$. At time $t=0$ both the inert spheres are situated at the same distance from $R$. (b) The colour plot for the concentration of B-type solvent particles.}
 	\label{fig:sch-conf}
 \end{figure}

Self-assembly of the active matter giving rise to complex architecture like living clusters due to symmetry breaking has been widely observed ~\cite{Singh:2017,Schmidt:2019,Gonzalez:2019,Bar:2006}. The phoretic interaction followed by short-range van der Waals and critical Casimir attractions are argued to be responsible for such clustering in case of the active particles~\cite{Singh:2017,Schmidt:2019}. To understand such self-assembly, here we probe the dynamics of two inert colloids $I_1, I_2$with radii $R_{i}=2.0$ in the presence of a reactive colloid $R$ with radius $R_{r}=4.0$. Fig.~\ref{fig:sch-conf}(a) shows the schematic representation of the initial configuration of the system where both the inert spheres are situated at the same distance from $R$ at time $t=0$. As discussed in the previous section, the $I_1$ and $I_2$ spheres will initially exhibit phoretic motion towards the reactive colloid, due to the concentration gradient of product, $B$ species around them as shown in Fig.~\ref{fig:sch-conf}(b). In this section, we are interested in motion of inert colloids after getting captured to the surface of $R$.

Before starting the discussion on what happens after capture, we would like to discuss the capture dynamics for the two inert colloid system. Table~\ref{tab_vI} below compares the velocity of the inert colloids at various $L_d$ as it approaches the reactive target sphere for the initial configuration given in Fig.~\ref{fig:sch-conf}. It is clear that both the inert spheres move with similar speed as they approach the target. This is because both of them experience the same diffusiophoretic pull by the $R$ sphere, by virtue of being at a similar distance from $R$.

\begin{table} [h!]
	\centering
	\begin{tabular}{ |p{1cm}|p{3.0cm}| p{3.0cm}| }
		\hline
		$L_{d}$ & $V_{I_1}$ & $V_{I_2}$\\  
		\hline 
		8.0 & 0.047 $\pm$ 0.007 & 0.040 $\pm$ 0.006\\
		\hline
		10.0 & 0.036 $\pm$ 0.010 & 0.032 $\pm$ 0.004\\
		\hline
		15.0& 0.018 $\pm$ 0.005 & 0.029 $\pm$ 0.009\\
		\hline
		20.0& 0.01$\pm$ 0.002 & 0.014 $\pm$ 0.003\\
		\hline
	\end{tabular}
	\caption{Propulsion velocity of $I_{1}$ and $I_{2}$ as a function of initial separation $L_{d}$ for configuration $C_1$.}
	\label{tab_vI}
\end{table}

\begin{figure} [h!]
	\centering
	\includegraphics[width=0.7\linewidth]{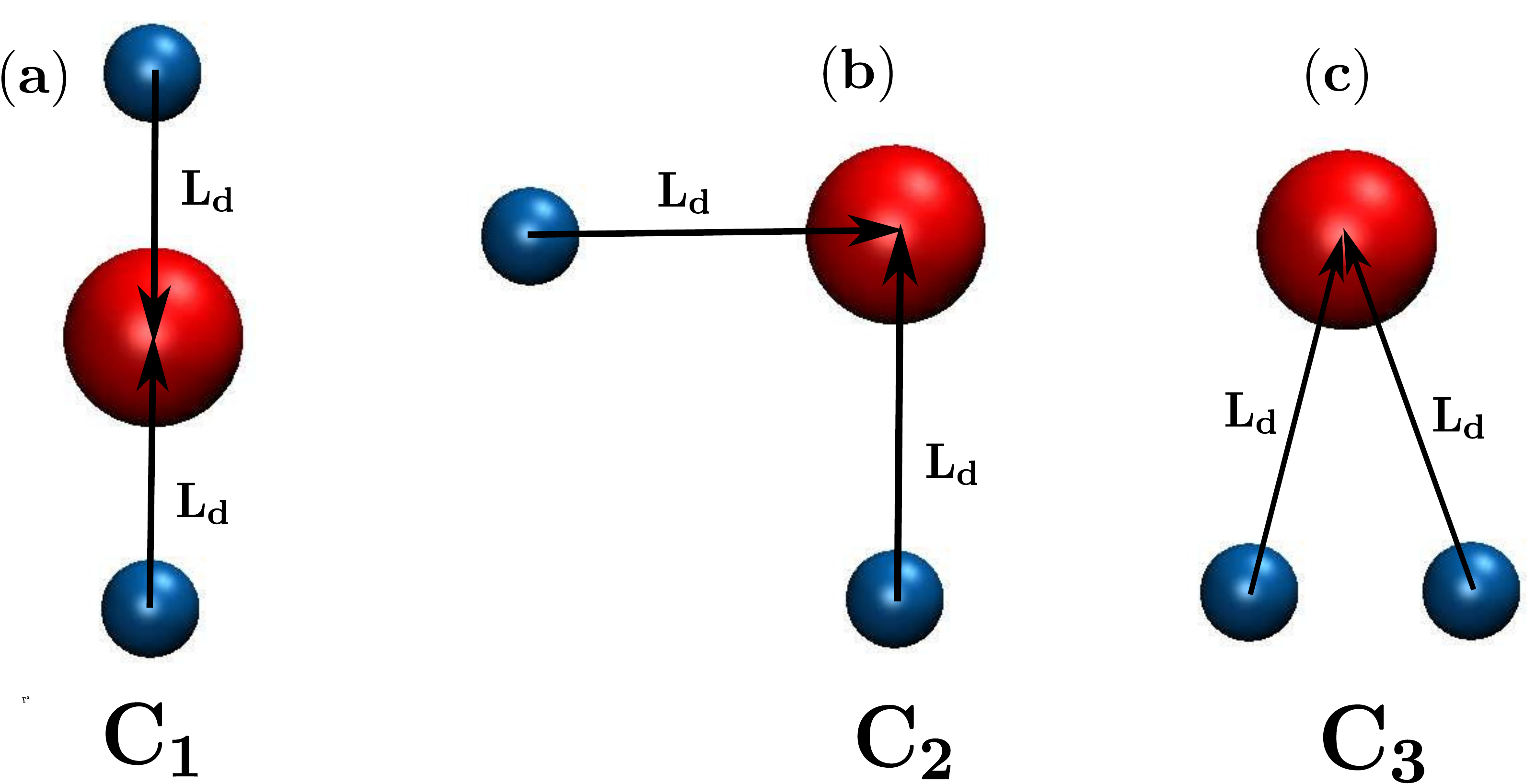}
	\caption{The initial configurations when two $I$ colloids approach the $R$ colloid at (a) $180{\degree}$  $(C_1)$ (b) $90{\degree}$ $(C_2)$ (c) $30{\degree}$ $(C_3)$.}
	\label{init_conf}
\end{figure}

\begin{figure} [h!]
	\centering
	\includegraphics[width=1.0\linewidth]{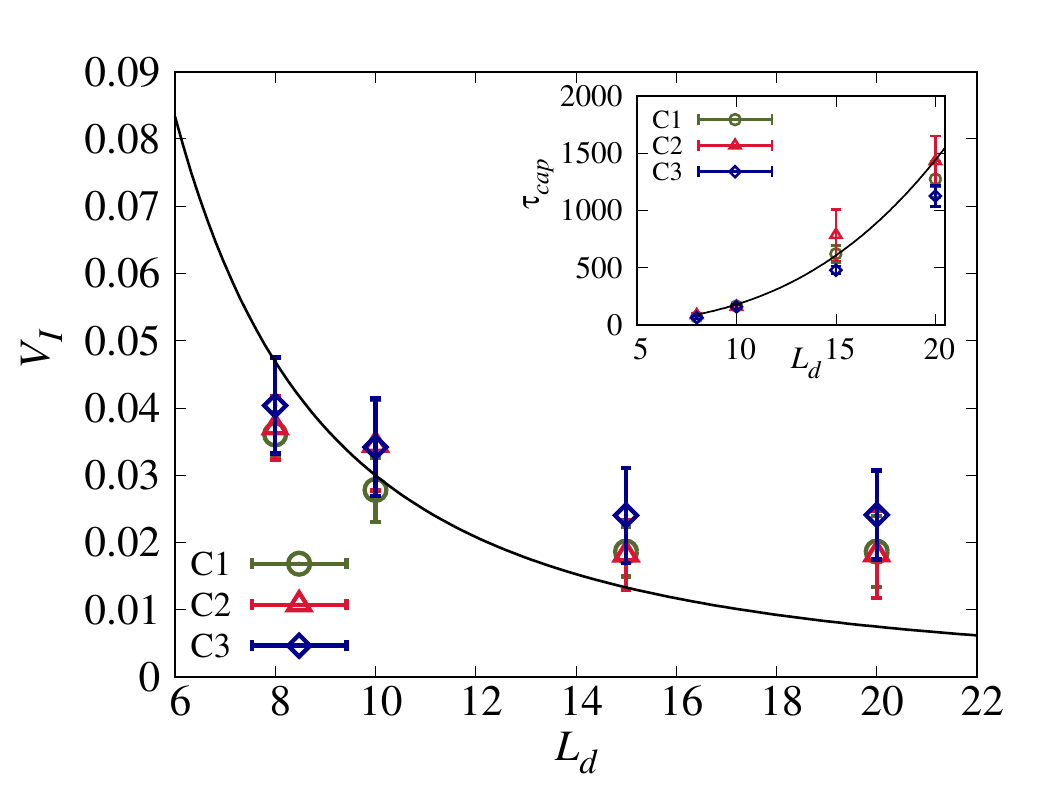}
	\caption{The average velocity $(V_{I})$ of $I$ colloids for different initial distances $(L_{d})$ from the reactive colloid for different initial configurations in the main plot and the average capture time ($\tau_{cap}$) in the inset.}
	\label{veln_tau}
\end{figure}

Further, to check the effect of the initial configuration of the system on the capture dynamics of inert spheres, we investigated three different initial arrangements. A representation of the initial set up is shown in Fig.\ref{init_conf}, where $C_{1}, C_{2}, C_{3}$ have colloidal arrangement such that the vectors joining $I$ and $R$ are at an angle $180{\degree}$, $90{\degree}$ and $30{\degree}$ respectively. As the reactive sphere $R$ sets up a radial gradient of product $B$, we expect the similar velocity and hence capture time of inert colloids as initially they are at the same radial distance from $R$. Fig.~\ref{veln_tau} compares the average directed velocity of inert colloids in these three configurations and concludes that the directed velocity of the inert sphere for the capture to happen depends only the on the radial distance between the target and walker.

\subsection{Spontaneous self-assembly by symmetry breaking }\label{sub2}

\begin{figure} [h!]
	\centering
	\includegraphics[width=1.0\linewidth]{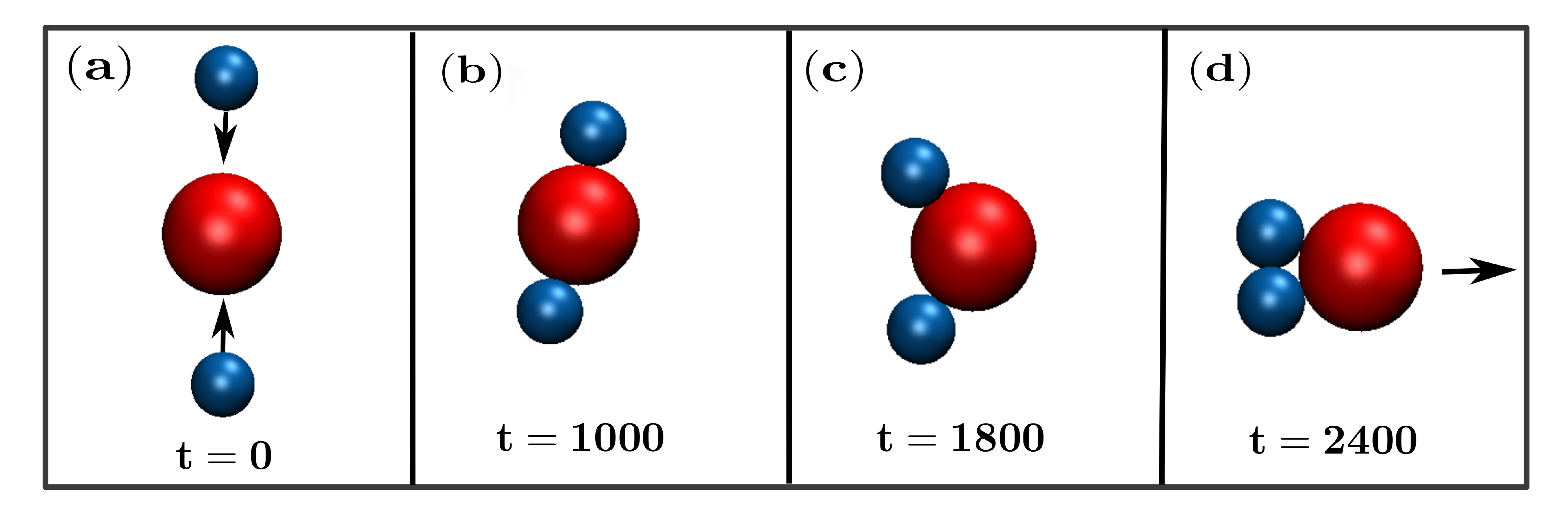}
	\includegraphics[width=0.97\linewidth]{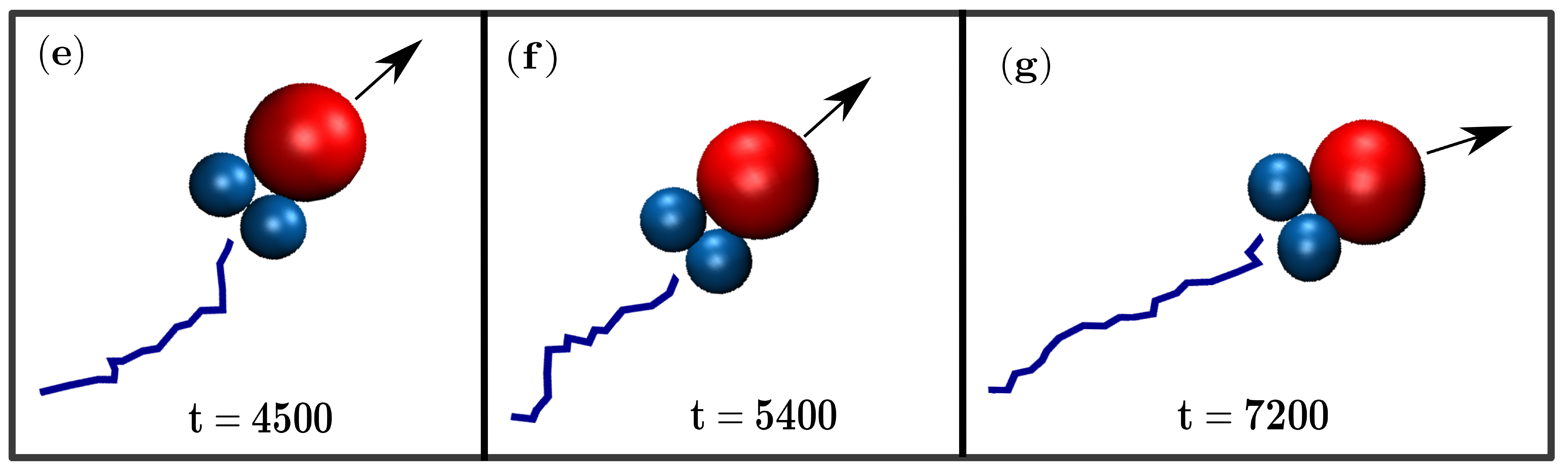}    
	\caption{Configurations depicting trimer formation which is ready to move in upper panel. The motion of trimer motor with represntative trajectories in lower panel, for the radii of reactive and inert colloids as $R_{r}=4.0$ and $R_{i}=2.0$ respectively. } 
	\label{conf}
\end{figure}

Once the colloids get captured to the reactive target sphere, we expect them to make a diffusive cluster if the capture of $I$s is head-on. This is because the head-on capture will lead to a symmetric configuration, which can not exhibit any directed motion. However, we observe a dramatic configurational adjustment leading to breaking of symmetry, which then induces directed propulsion to the cluster. During this adjustment, both the inert spheres self-assemble on one side of the reactive sphere, to obtain a configuration similar to Janus dimer in the previous subsection~\ref{sub1}. Fig.~\ref{conf} shows the process of capture and self-assembly of inert spheres on $R$. It is important to note that the self-assembly of $I$s on $R$ which then leads to spontaneous motor formation is independent of initial configuration and $C_2$, $C_3$ also exhibit a similar dynamics.

Another noteworthy observation, in this case, is about the center-of-mass speed ($V_{cm}$) of the propelling aggregate. We compare the $V_{cm}$ of the cluster for single $I$ and a pair of $I$ case and report it in Table~\ref{tab}. It is evident that in the presence of more $I$ the speed of cluster increases as the force-generating entities have increased.

\begin{table} [h!]
	\centering
	\begin{tabular}{|p{1.2cm}||p{3.5cm}|}
		\hline
		$V_{cm}^{1I}$ & 0.0390 $\pm$ 0.0001\\
		\hline
		$V_{cm}^{2I}$& 0.0451 $\pm$ 0.0001\\
		\hline
	\end{tabular}
	\caption{Speed of motor shows increment when we deal with two inert colloidal system in the presence of one reactive colloid.}
	\label{tab}
\end{table}

The obvious question that arises from the above observation is, why do the inert particles cluster on one side of the reactive sphere, and how does it chooses the direction to form such of assembly? To answer these queries, we calculate the time average force experienced by the inert spheres due to solvent as a function of  azimuthal angle (defined in Fig.~\ref{fig:sch-conf}(a)) between configurations like Fig.~\ref{conf}(b) and (c), \textit{i.e.} time span after the capture and before the assembly. For a particular ensemble shown in Fig.~\ref{conf}, the force calculation depicts higher force on right sides of $I$ spheres; therefore, they must move towards the left hemisphere of $R$ and indeed we observe them moving in the same direction.

\begin{figure} [!h]
	\centering
	\includegraphics[width=0.9\linewidth]{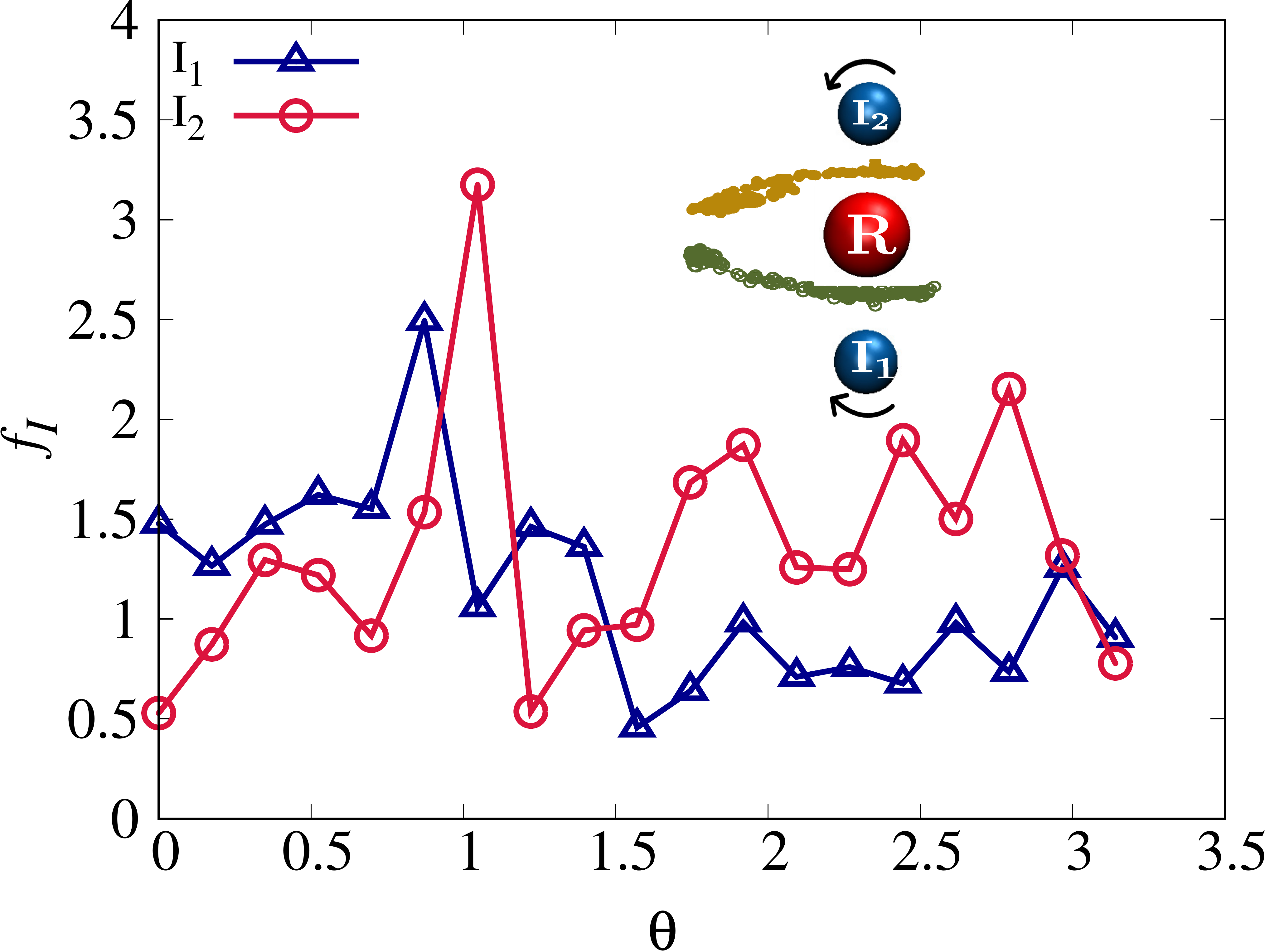}    
	\caption{ The force exerted by the fluid particles on the inert colloids during self-assembly process as function of azimuthal angle $\theta$. Here, $Is$ assemble towards the left hemi-sphere of $R$.} 
	\label{forces}
\end{figure} 

\begin{figure} [!h]
	\centering
	\includegraphics[width=1.0\linewidth]{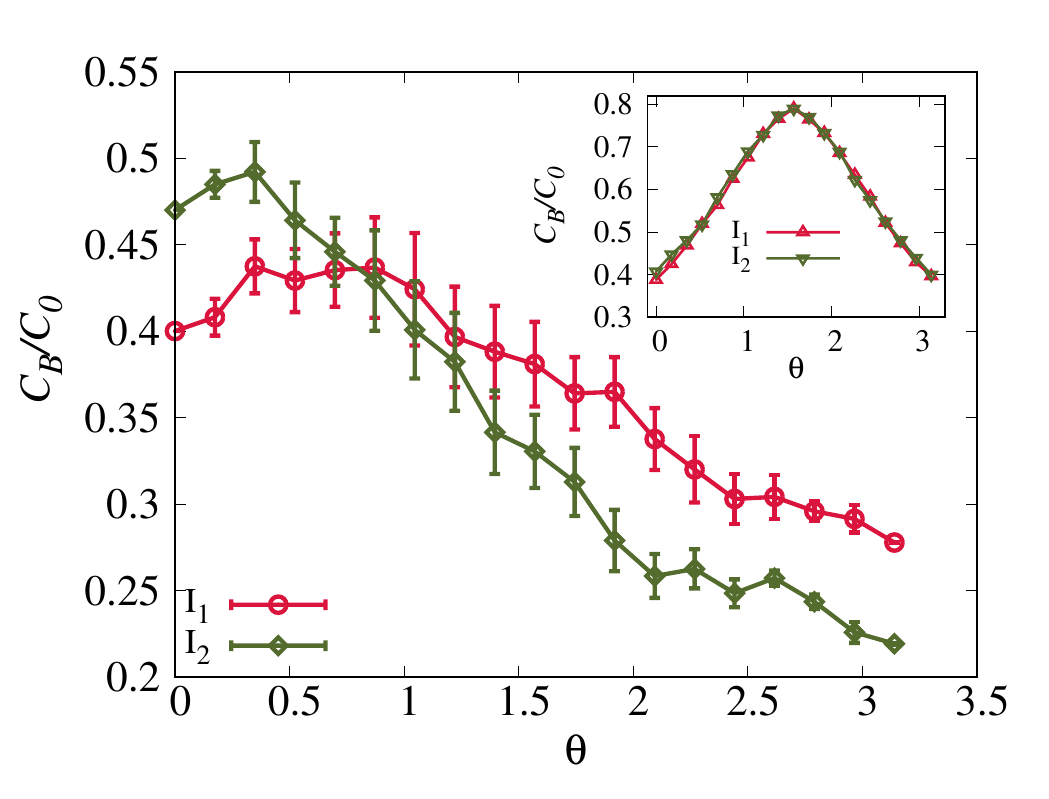}
		\caption{The concentration of $B$ fluid particles around $I_{1}$ and $I_{2}$ respectively when they have started assembling towards the left hemisphere of $R$ colloid whereas the inset shows concentration of $B$ when $I_{1}$ and $I_{2}$ are far from $R$ at $t=0$.} 
	\label{B_dis}
\end{figure}

Of course, the next question to ask would be what creates such force asymmetry on inert spheres. The diffusiophoretic model tells us that any chemical gradient is sufficient to provide the force asymmetry. Therefore, we calculate the concentration of $B$ type fluid particles around the inert colloids. The normalised concentration of $B$ fluid particles $(C_{B}/C_{0})$ as a function of $\theta$, where $\theta$ is defined in Fig.~\ref{fig:sch-conf}(a) is plotted in Fig.~\ref{B_dis}. This clearly depicts the presence of chemical gradient around the inert particles, which then leads to force asymmetry and hence self-assembly of $I$ colloids on $R$.

Fig.\ref{B_dis}(inset) shows the normalised concentration of $B$ fluid particles $(C_{B}/C_{0})$  as a function of $\theta$ when the $I$s are not captured on $R$, \textit{i.e.} they are sufficiently far from the target. Hence when $I$s are sufficiently away from $R$, there is no bias on them to choose the direction of assembly. The choice of the direction of the assembly shown by inert colloids after capture is completely arbitrary, which depends on the very local asymmetry of forces and fluid distribution.

\subsection{Factors affecting the self-assembly}\label{sub3}

Self-assembly of inert spheres on $R$ is influenced by factors like the interaction of solvent with $I$, hydrodynamic interactions in the system, the interaction strength, etc. We probe some of them below, and the quantification is done on the basis of time taken by $I$ spheres to self-assemble on $R$ ($\tau_{assm}$) after getting captured on its surface. 

\begin{figure} [h!]
	\centering
	\includegraphics[width=1.0\linewidth]{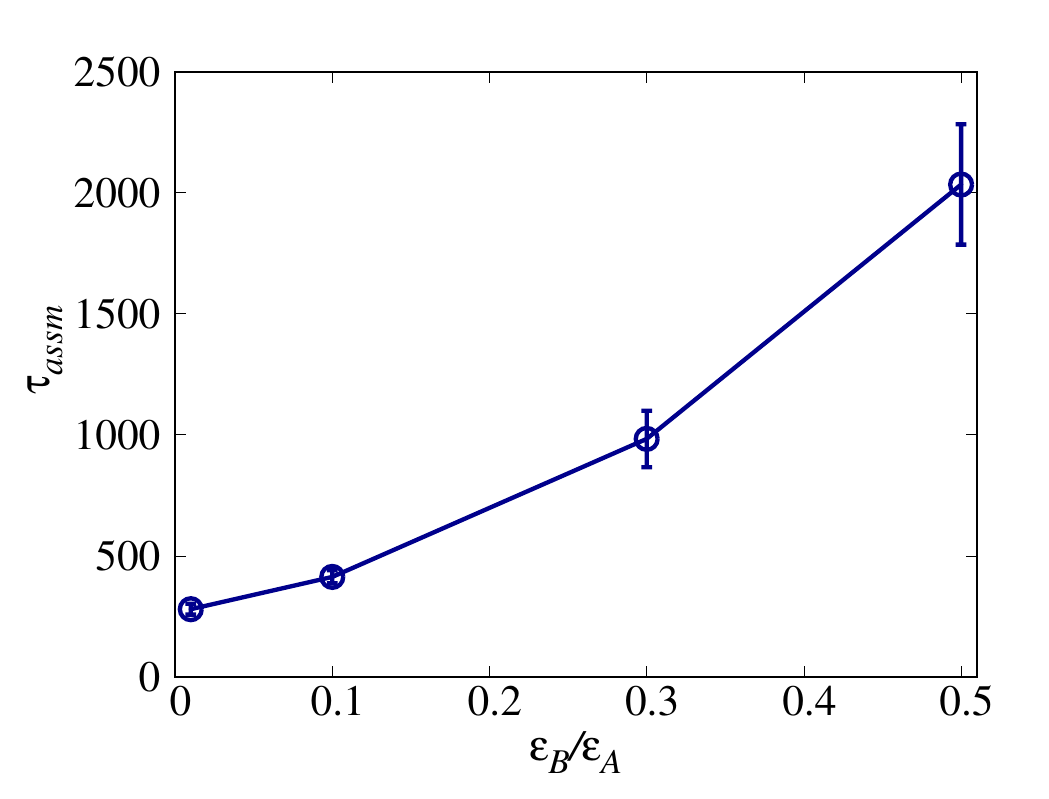}
	\caption{The average time taken by $I$ colloids to self-assemble after their capture at $R$ which is varying with $\epsilon_{B}$.}
	\label{ass_eb}
\end{figure}

For self-assembly point of view, we believe that the interactions of solvent with $I$ is a critical parameter. The ratio of $\epsilon_{B}/\epsilon_{A}$ decides the asymmetric force experienced by $I$, which in turns decides its motility. For $\epsilon_{B}/\epsilon_{A}=1.0$ the inert particles are purely diffusive and do not show any propulsion and assembly. Therefore, to probe the effect of solvent interaction on self-assembly, we plot, $\tau_{assm}$ vs.  $\epsilon_{B}/\epsilon_{A}$ in Fig.~\ref{ass_eb}. For very small values of $\epsilon_{B}/\epsilon_{A}$ \textit{i.e.} for more asymmetry in forces, self-assembly is more rapid leading to smaller $\tau_{assm}$ supporting the fact that local fluid interaction with $I$ is one of the deciding factors for self-assembly on the target sphere.

The hydrodynamic interactions (HI) involved during the capture and self-assembly process has its unique role~\cite{Marchetti:2013,Singh:2016,Shen:2019}. To confirm the role of the hydrodynamic effect further, we calculate $\tau_{assm}$ for the 2$I$ system by switching off HI. The advantage of MPCD simulation lies in the fact that it is possible to switch off the HI effect whenever needed. The standard method to suppress HI in MPC simulations is to randomize the velocities of the solvents by re-sampling the velocities from the Maxwell-Boltzmann distribution at every collision step. In this case, we do observe a similar capture and assembly process; however, the speed of the capture and self-assembly slows down to a very large extent.

\subsection{Assembly of multiple inert colloids leading to large motile cluster}

\begin{figure} [h!]
	\centering
	\includegraphics[width=1.0\linewidth]{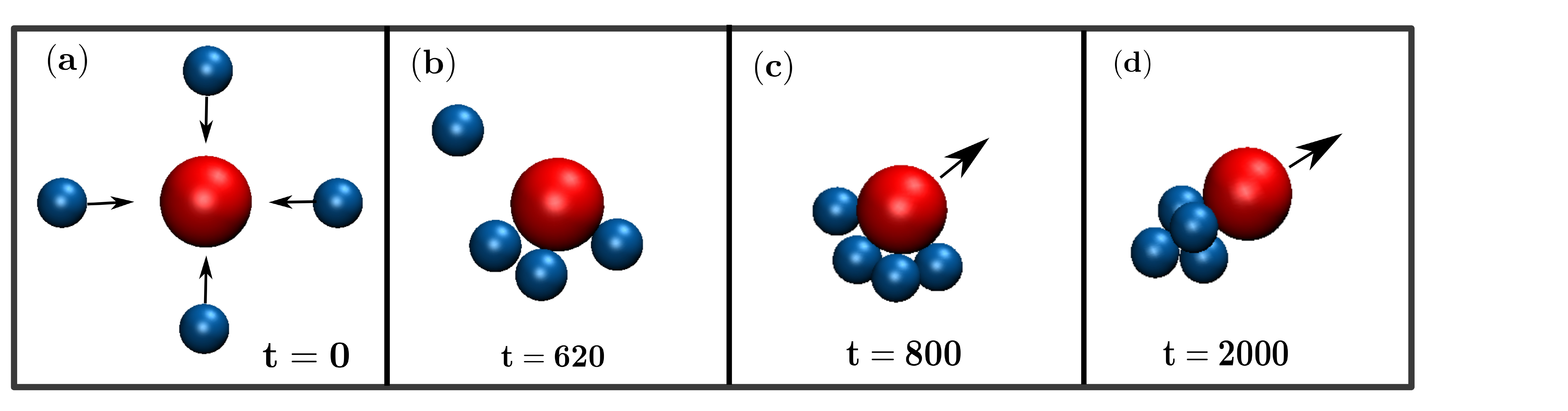}
	\caption{Formation of a pentamer with four $I$ and one $R$ system which is moving with a constant speed. Here $R_{i}=2.0$ and $R_{r}=4.0$}
	\label{4I_1R}
\end{figure}

The work by Schmidt \textit{et al.} on active self-assembly induced by light in a suspension colloid has demonstrated the dynamic formation of complex architectures such as living clusters~\cite{Schmidt:2019}. Therefore, it would definitely be interesting to ask whether such dynamic self-assembly into a plethora of complex structure is possible in a chemical system like ours. To answer this question, we add multiple inert colloids in the system keeping a single $R$ sphere. In this case also the $I$ spheres get captured on the surface of diffusive $R$ colloid, which then self-assemble on one side of the surface and converts the assembly on a self-propelled cluster. An example of such a case is shown in Fig.~\ref{4I_1R} where a system comprises of four $I$s and one $R$ shows the formation of an assembled structure (Fig.~\ref{4I_1R}(b)) which then can self-propel as a dynamical motor (Fig.~\ref{4I_1R}(c) and (d)).

Therefore the above observed chemically powered self-assembly leading to the formation of the chemical motor is quite robust and provides an easy route to obtain large chemical motors from a dispersion of spherically symmetric colloids.

\section{Conclusions}\label{conclusion}

In conclusion, we have presented a simple strategy for the formation of self-propelling chemical motors via the process of self-assembly. In this work, we study a chemical system which comprises a chemically reactive colloid that converts the fuel to the product, which is then sensed by inert colloids in its vicinity. Two species of isotropic colloids that cannot self-propel by themselves interact to form a self-assembled cluster that can swim. Here, we identify and discuss the factors that are responsible for such self-assembly and self-propulsion. The process of assembly and propulsion is highly scalable, and we show the formation of dimers to pentamers in our study. 

We show that the chemical field around the reactive colloid causes the binding of inert spheres on the surface of reactive one through diffusiophoresis. The assembly then forms a cluster which by virtue of symmetry breaking gets converted to a self-propelled cluster.  Further, the capture and assembly process of passive colloids is independent of the initial configuration. The capture time, as well as the propulsion velocity of inert colloids, depends on its initial distance from active colloid and its interaction with the fluid. 

The aggregation of colloidal particles is an outcome of dynamical self-organization, leading to the formation of a motile motor is an interesting study. Such self-motile motors are involved in many biological processes also; for example, the self-assembly process of kinesin motor proteins have implication in formation of the mitotic spindle~\cite{Sawin:1992}.

\acknowledgements The computational work was carried out at the HPC facility in IISER Bhopal, India. ST acknowledges SERB, DST (grant no. EMR/2017/000922) for funding.

 \bibliography{reference.bib}

\begin{thebibliography}{36}%
\makeatletter
\providecommand \@ifxundefined [1]{%
 \@ifx{#1\undefined}
}%
\providecommand \@ifnum [1]{%
 \ifnum #1\expandafter \@firstoftwo
 \else \expandafter \@secondoftwo
 \fi
}%
\providecommand \@ifx [1]{%
 \ifx #1\expandafter \@firstoftwo
 \else \expandafter \@secondoftwo
 \fi
}%
\providecommand \natexlab [1]{#1}%
\providecommand \enquote  [1]{``#1''}%
\providecommand \bibnamefont  [1]{#1}%
\providecommand \bibfnamefont [1]{#1}%
\providecommand \citenamefont [1]{#1}%
\providecommand \href@noop [0]{\@secondoftwo}%
\providecommand \href [0]{\begingroup \@sanitize@url \@href}%
\providecommand \@href[1]{\@@startlink{#1}\@@href}%
\providecommand \@@href[1]{\endgroup#1\@@endlink}%
\providecommand \@sanitize@url [0]{\catcode `\\12\catcode `\$12\catcode
  `\&12\catcode `\#12\catcode `\^12\catcode `\_12\catcode `\%12\relax}%
\providecommand \@@startlink[1]{}%
\providecommand \@@endlink[0]{}%
\providecommand \url  [0]{\begingroup\@sanitize@url \@url }%
\providecommand \@url [1]{\endgroup\@href {#1}{\urlprefix }}%
\providecommand \urlprefix  [0]{URL }%
\providecommand \Eprint [0]{\href }%
\providecommand \doibase [0]{http://dx.doi.org/}%
\providecommand \selectlanguage [0]{\@gobble}%
\providecommand \bibinfo  [0]{\@secondoftwo}%
\providecommand \bibfield  [0]{\@secondoftwo}%
\providecommand \translation [1]{[#1]}%
\providecommand \BibitemOpen [0]{}%
\providecommand \bibitemStop [0]{}%
\providecommand \bibitemNoStop [0]{.\EOS\space}%
\providecommand \EOS [0]{\spacefactor3000\relax}%
\providecommand \BibitemShut  [1]{\csname bibitem#1\endcsname}%
\let\auto@bib@innerbib\@empty
%</preamble>
\bibitem [{\citenamefont {Zhang}\ \emph {et~al.}(2010)\citenamefont {Zhang},
  \citenamefont {E.-L.},\ and\ \citenamefont {Swinney}}]{Zhang:2010}%
  \BibitemOpen
  \bibfield  {author} {\bibinfo {author} {\bibfnamefont {A.~F.}\ \bibnamefont
  {Zhang}, \bibfnamefont {H.~P.and~Be'er}}, \bibinfo {author} {\bibnamefont
  {E.-L.}}, \ and\ \bibinfo {author} {\bibfnamefont {H.~L.}\ \bibnamefont
  {Swinney}},\ }\href@noop {} {\bibfield  {journal} {\bibinfo  {journal} {Proc
  Natl Acad Sci U S A}\ }\textbf {\bibinfo {volume} {107}},\ \bibinfo {pages}
  {13626} (\bibinfo {year} {2010})}\BibitemShut {NoStop}%
\bibitem [{\citenamefont {Eisenbach}\ and\ \citenamefont
  {Giojalas}(2006)}]{Eisenbach:2006}%
  \BibitemOpen
  \bibfield  {author} {\bibinfo {author} {\bibfnamefont {M.}~\bibnamefont
  {Eisenbach}}\ and\ \bibinfo {author} {\bibfnamefont {L.~C.}\ \bibnamefont
  {Giojalas}},\ }\href@noop {} {\bibfield  {journal} {\bibinfo  {journal} {Nat.
  Rev. Mol. Cell Biol.}\ }\textbf {\bibinfo {volume} {7}},\ \bibinfo {pages}
  {276} (\bibinfo {year} {2006})}\BibitemShut {NoStop}%
\bibitem [{\citenamefont {Li}\ \emph {et~al.}(2005)\citenamefont {Li},
  \citenamefont {Tang}, \citenamefont {Armes}, \citenamefont {Morris},
  \citenamefont {Rose}, \citenamefont {Lloyd},\ and\ \citenamefont
  {Lewis}}]{Li:2005}%
  \BibitemOpen
  \bibfield  {author} {\bibinfo {author} {\bibfnamefont {C.}~\bibnamefont
  {Li}}, \bibinfo {author} {\bibfnamefont {Y.}~\bibnamefont {Tang}}, \bibinfo
  {author} {\bibfnamefont {S.~P.}\ \bibnamefont {Armes}}, \bibinfo {author}
  {\bibfnamefont {C.~J.}\ \bibnamefont {Morris}}, \bibinfo {author}
  {\bibfnamefont {S.~F.}\ \bibnamefont {Rose}}, \bibinfo {author}
  {\bibfnamefont {A.~W.}\ \bibnamefont {Lloyd}}, \ and\ \bibinfo {author}
  {\bibfnamefont {A.~L.}\ \bibnamefont {Lewis}},\ }\href@noop {} {\bibfield
  {journal} {\bibinfo  {journal} {Biomacromolecules}\ }\textbf {\bibinfo
  {volume} {6}},\ \bibinfo {pages} {994} (\bibinfo {year} {2005})}\BibitemShut
  {NoStop}%
\bibitem [{\citenamefont {Gao}\ \emph {et~al.}(2013)\citenamefont {Gao},
  \citenamefont {Pei}, \citenamefont {Feng}, \citenamefont {Hennessy},\ and\
  \citenamefont {Wang}}]{Gao:2013}%
  \BibitemOpen
  \bibfield  {author} {\bibinfo {author} {\bibfnamefont {W.}~\bibnamefont
  {Gao}}, \bibinfo {author} {\bibfnamefont {A.}~\bibnamefont {Pei}}, \bibinfo
  {author} {\bibfnamefont {X.}~\bibnamefont {Feng}}, \bibinfo {author}
  {\bibfnamefont {C.}~\bibnamefont {Hennessy}}, \ and\ \bibinfo {author}
  {\bibfnamefont {J.}~\bibnamefont {Wang}},\ }\href@noop {} {\bibfield
  {journal} {\bibinfo  {journal} {J. Am. Chem. Soc.}\ }\textbf {\bibinfo
  {volume} {135}},\ \bibinfo {pages} {998} (\bibinfo {year}
  {2013})}\BibitemShut {NoStop}%
\bibitem [{\citenamefont {Sundararajan}\ \emph {et~al.}(2008)\citenamefont
  {Sundararajan}, \citenamefont {Lammert}, \citenamefont {Zudans},
  \citenamefont {Crespi},\ and\ \citenamefont {Sen}}]{Sundararajan:2008}%
  \BibitemOpen
  \bibfield  {author} {\bibinfo {author} {\bibfnamefont {S.}~\bibnamefont
  {Sundararajan}}, \bibinfo {author} {\bibfnamefont {P.~E.}\ \bibnamefont
  {Lammert}}, \bibinfo {author} {\bibfnamefont {A.~W.}\ \bibnamefont {Zudans}},
  \bibinfo {author} {\bibfnamefont {V.~H.}\ \bibnamefont {Crespi}}, \ and\
  \bibinfo {author} {\bibfnamefont {A.}~\bibnamefont {Sen}},\ }\href@noop {}
  {\bibfield  {journal} {\bibinfo  {journal} {Nano Lett.}\ }\textbf {\bibinfo
  {volume} {8}},\ \bibinfo {pages} {1271} (\bibinfo {year} {2008})}\BibitemShut
  {NoStop}%
\bibitem [{\citenamefont {Guix}\ \emph {et~al.}(2014)\citenamefont {Guix},
  \citenamefont {Mayorga-Martinez},\ and\ \citenamefont
  {Merko{\c{c}}i}}]{Guix:2014}%
  \BibitemOpen
  \bibfield  {author} {\bibinfo {author} {\bibfnamefont {M.}~\bibnamefont
  {Guix}}, \bibinfo {author} {\bibfnamefont {C.~C.}\ \bibnamefont
  {Mayorga-Martinez}}, \ and\ \bibinfo {author} {\bibfnamefont
  {A.}~\bibnamefont {Merko{\c{c}}i}},\ }\href@noop {} {\bibfield  {journal}
  {\bibinfo  {journal} {Chem. Rev.}\ }\textbf {\bibinfo {volume} {114}},\
  \bibinfo {pages} {6285} (\bibinfo {year} {2014})}\BibitemShut {NoStop}%
\bibitem [{\citenamefont {Yu}\ \emph {et~al.}(2018)\citenamefont {Yu},
  \citenamefont {Chuphal}, \citenamefont {Thakur}, \citenamefont {Reigh},
  \citenamefont {Singh},\ and\ \citenamefont {Fischer}}]{Yu:2018}%
  \BibitemOpen
  \bibfield  {author} {\bibinfo {author} {\bibfnamefont {T.}~\bibnamefont
  {Yu}}, \bibinfo {author} {\bibfnamefont {P.}~\bibnamefont {Chuphal}},
  \bibinfo {author} {\bibfnamefont {S.}~\bibnamefont {Thakur}}, \bibinfo
  {author} {\bibfnamefont {S.~Y.}\ \bibnamefont {Reigh}}, \bibinfo {author}
  {\bibfnamefont {D.~P.}\ \bibnamefont {Singh}}, \ and\ \bibinfo {author}
  {\bibfnamefont {P.}~\bibnamefont {Fischer}},\ }\href@noop {} {\bibfield
  {journal} {\bibinfo  {journal} {Chem. Commun.}\ }\textbf {\bibinfo {volume}
  {54}},\ \bibinfo {pages} {11933} (\bibinfo {year} {2018})}\BibitemShut
  {NoStop}%
\bibitem [{\citenamefont {Reigh}\ \emph {et~al.}(2018)\citenamefont {Reigh},
  \citenamefont {Chuphal}, \citenamefont {Thakur},\ and\ \citenamefont
  {Kapral}}]{SY:2018}%
  \BibitemOpen
  \bibfield  {author} {\bibinfo {author} {\bibfnamefont {S.~Y.}\ \bibnamefont
  {Reigh}}, \bibinfo {author} {\bibfnamefont {P.}~\bibnamefont {Chuphal}},
  \bibinfo {author} {\bibfnamefont {S.}~\bibnamefont {Thakur}}, \ and\ \bibinfo
  {author} {\bibfnamefont {R.}~\bibnamefont {Kapral}},\ }\href@noop {}
  {\bibfield  {journal} {\bibinfo  {journal} {Soft Matter}\ }\textbf {\bibinfo
  {volume} {14}},\ \bibinfo {pages} {6043} (\bibinfo {year}
  {2018})}\BibitemShut {NoStop}%
\bibitem [{\citenamefont {Palacci}\ \emph {et~al.}(2013)\citenamefont
  {Palacci}, \citenamefont {Sacanna}, \citenamefont {Steinberg}, \citenamefont
  {Pine},\ and\ \citenamefont {Chaikin}}]{Palacci:2013}%
  \BibitemOpen
  \bibfield  {author} {\bibinfo {author} {\bibfnamefont {J.}~\bibnamefont
  {Palacci}}, \bibinfo {author} {\bibfnamefont {S.}~\bibnamefont {Sacanna}},
  \bibinfo {author} {\bibfnamefont {A.~P.}\ \bibnamefont {Steinberg}}, \bibinfo
  {author} {\bibfnamefont {D.~J.}\ \bibnamefont {Pine}}, \ and\ \bibinfo
  {author} {\bibfnamefont {P.~M.}\ \bibnamefont {Chaikin}},\ }\href@noop {}
  {\bibfield  {journal} {\bibinfo  {journal} {Science}\ }\textbf {\bibinfo
  {volume} {339}},\ \bibinfo {pages} {936} (\bibinfo {year}
  {2013})}\BibitemShut {NoStop}%
\bibitem [{\citenamefont {Schmidt}\ \emph {et~al.}(2019)\citenamefont
  {Schmidt}, \citenamefont {Liebchen}, \citenamefont {Löwen},\ and\
  \citenamefont {Volpe}}]{Schmidt:2019}%
  \BibitemOpen
  \bibfield  {author} {\bibinfo {author} {\bibfnamefont {F.}~\bibnamefont
  {Schmidt}}, \bibinfo {author} {\bibfnamefont {B.}~\bibnamefont {Liebchen}},
  \bibinfo {author} {\bibfnamefont {H.}~\bibnamefont {Löwen}}, \ and\ \bibinfo
  {author} {\bibfnamefont {G.}~\bibnamefont {Volpe}},\ }\href@noop {}
  {\bibfield  {journal} {\bibinfo  {journal} {J. Chem. Phys.}\ }\textbf
  {\bibinfo {volume} {150}},\ \bibinfo {pages} {094905} (\bibinfo {year}
  {2019})}\BibitemShut {NoStop}%
\bibitem [{\citenamefont {Gong}\ \emph {et~al.}(2019)\citenamefont {Gong},
  \citenamefont {Sun}, \citenamefont {Zhang}, \citenamefont {Sun},
  \citenamefont {Su}, \citenamefont {Wu},\ and\ \citenamefont
  {Wei}}]{Gong:2019}%
  \BibitemOpen
  \bibfield  {author} {\bibinfo {author} {\bibfnamefont {C.}~\bibnamefont
  {Gong}}, \bibinfo {author} {\bibfnamefont {S.}~\bibnamefont {Sun}}, \bibinfo
  {author} {\bibfnamefont {Y.}~\bibnamefont {Zhang}}, \bibinfo {author}
  {\bibfnamefont {L.}~\bibnamefont {Sun}}, \bibinfo {author} {\bibfnamefont
  {Z.}~\bibnamefont {Su}}, \bibinfo {author} {\bibfnamefont {A.}~\bibnamefont
  {Wu}}, \ and\ \bibinfo {author} {\bibfnamefont {G.}~\bibnamefont {Wei}},\
  }\href@noop {} {\bibfield  {journal} {\bibinfo  {journal} {Nanoscale}\
  }\textbf {\bibinfo {volume} {11}},\ \bibinfo {pages} {4147} (\bibinfo {year}
  {2019})}\BibitemShut {NoStop}%
\bibitem [{\citenamefont {Oerlemans}\ \emph {et~al.}(2010)\citenamefont
  {Oerlemans}, \citenamefont {Bult}, \citenamefont {Bos}, \citenamefont
  {Storm}, \citenamefont {Nijsen},\ and\ \citenamefont
  {Hennink}}]{Oerlemans:2010}%
  \BibitemOpen
  \bibfield  {author} {\bibinfo {author} {\bibfnamefont {C.}~\bibnamefont
  {Oerlemans}}, \bibinfo {author} {\bibfnamefont {W.}~\bibnamefont {Bult}},
  \bibinfo {author} {\bibfnamefont {M.}~\bibnamefont {Bos}}, \bibinfo {author}
  {\bibfnamefont {G.}~\bibnamefont {Storm}}, \bibinfo {author} {\bibfnamefont
  {J.~F.~W.}\ \bibnamefont {Nijsen}}, \ and\ \bibinfo {author} {\bibfnamefont
  {W.~E.}\ \bibnamefont {Hennink}},\ }\href@noop {} {\bibfield  {journal}
  {\bibinfo  {journal} {Pharm. Res.}\ }\textbf {\bibinfo {volume} {27}},\
  \bibinfo {pages} {2569} (\bibinfo {year} {2010})}\BibitemShut {NoStop}%
\bibitem [{\citenamefont {Alapan}\ \emph {et~al.}(2018)\citenamefont {Alapan},
  \citenamefont {Yasa}, \citenamefont {Schauer}, \citenamefont {Giltinan},
  \citenamefont {Tabak}, \citenamefont {Sourjik},\ and\ \citenamefont
  {Sitti}}]{Alapan:2018}%
  \BibitemOpen
  \bibfield  {author} {\bibinfo {author} {\bibfnamefont {Y.}~\bibnamefont
  {Alapan}}, \bibinfo {author} {\bibfnamefont {O.}~\bibnamefont {Yasa}},
  \bibinfo {author} {\bibfnamefont {O.}~\bibnamefont {Schauer}}, \bibinfo
  {author} {\bibfnamefont {J.}~\bibnamefont {Giltinan}}, \bibinfo {author}
  {\bibfnamefont {A.~F.}\ \bibnamefont {Tabak}}, \bibinfo {author}
  {\bibfnamefont {V.}~\bibnamefont {Sourjik}}, \ and\ \bibinfo {author}
  {\bibfnamefont {M.}~\bibnamefont {Sitti}},\ }\href@noop {} {\bibfield
  {journal} {\bibinfo  {journal} {Sci. Robot.}\ }\textbf {\bibinfo {volume}
  {3}},\ \bibinfo {pages} {eaar4423} (\bibinfo {year} {2018})}\BibitemShut
  {NoStop}%
\bibitem [{\citenamefont {Yan}\ \emph {et~al.}(2012)\citenamefont {Yan},
  \citenamefont {Bloom}, \citenamefont {Bae}, \citenamefont {Luijten},\ and\
  \citenamefont {Granick}}]{Yan:2012}%
  \BibitemOpen
  \bibfield  {author} {\bibinfo {author} {\bibfnamefont {J.}~\bibnamefont
  {Yan}}, \bibinfo {author} {\bibfnamefont {M.}~\bibnamefont {Bloom}}, \bibinfo
  {author} {\bibfnamefont {S.~C.}\ \bibnamefont {Bae}}, \bibinfo {author}
  {\bibfnamefont {E.}~\bibnamefont {Luijten}}, \ and\ \bibinfo {author}
  {\bibfnamefont {S.}~\bibnamefont {Granick}},\ }\href@noop {} {\bibfield
  {journal} {\bibinfo  {journal} {Nature}\ }\textbf {\bibinfo {volume} {491}},\
  \bibinfo {pages} {578 EP } (\bibinfo {year} {2012})}\BibitemShut {NoStop}%
\bibitem [{\citenamefont {Ma}\ \emph {et~al.}(2017)\citenamefont {Ma},
  \citenamefont {Lei},\ and\ \citenamefont {Ni}}]{Ma:2017}%
  \BibitemOpen
  \bibfield  {author} {\bibinfo {author} {\bibfnamefont {Z.}~\bibnamefont
  {Ma}}, \bibinfo {author} {\bibfnamefont {Q.-l.}\ \bibnamefont {Lei}}, \ and\
  \bibinfo {author} {\bibfnamefont {R.}~\bibnamefont {Ni}},\ }\href@noop {}
  {\bibfield  {journal} {\bibinfo  {journal} {Soft Matter}\ }\textbf {\bibinfo
  {volume} {13}},\ \bibinfo {pages} {8940} (\bibinfo {year}
  {2017})}\BibitemShut {NoStop}%
\bibitem [{\citenamefont {Martin}\ and\ \citenamefont
  {Snezhko}(2013)}]{Martin:2013}%
  \BibitemOpen
  \bibfield  {author} {\bibinfo {author} {\bibfnamefont {J.~E.}\ \bibnamefont
  {Martin}}\ and\ \bibinfo {author} {\bibfnamefont {A.}~\bibnamefont
  {Snezhko}},\ }\href@noop {} {\bibfield  {journal} {\bibinfo  {journal} {Rep.
  Prog. Phys.}\ }\textbf {\bibinfo {volume} {76}},\ \bibinfo {pages} {126601}
  (\bibinfo {year} {2013})}\BibitemShut {NoStop}%
\bibitem [{\citenamefont {Takahara}\ \emph {et~al.}(2005)\citenamefont
  {Takahara}, \citenamefont {Ikeda}, \citenamefont {Ishino}, \citenamefont
  {Tachi}, \citenamefont {Ikeue}, \citenamefont {Sakata}, \citenamefont
  {Hasegawa}, \citenamefont {Mori}, \citenamefont {Matsumura},\ and\
  \citenamefont {Ohtani}}]{Takahara:2005}%
  \BibitemOpen
  \bibfield  {author} {\bibinfo {author} {\bibfnamefont {Y.~K.}\ \bibnamefont
  {Takahara}}, \bibinfo {author} {\bibfnamefont {S.}~\bibnamefont {Ikeda}},
  \bibinfo {author} {\bibfnamefont {S.}~\bibnamefont {Ishino}}, \bibinfo
  {author} {\bibfnamefont {K.}~\bibnamefont {Tachi}}, \bibinfo {author}
  {\bibfnamefont {K.}~\bibnamefont {Ikeue}}, \bibinfo {author} {\bibfnamefont
  {T.}~\bibnamefont {Sakata}}, \bibinfo {author} {\bibfnamefont
  {T.}~\bibnamefont {Hasegawa}}, \bibinfo {author} {\bibfnamefont
  {H.}~\bibnamefont {Mori}}, \bibinfo {author} {\bibfnamefont {M.}~\bibnamefont
  {Matsumura}}, \ and\ \bibinfo {author} {\bibfnamefont {B.}~\bibnamefont
  {Ohtani}},\ }\href@noop {} {\bibfield  {journal} {\bibinfo  {journal} {J. Am.
  Chem. Soc.}\ }\textbf {\bibinfo {volume} {127}},\ \bibinfo {pages} {6271}
  (\bibinfo {year} {2005})}\BibitemShut {NoStop}%
\bibitem [{\citenamefont {Bogart}\ \emph {et~al.}(2014)\citenamefont {Bogart},
  \citenamefont {Pourroy}, \citenamefont {Murphy}, \citenamefont {Puntes},
  \citenamefont {Pellegrino}, \citenamefont {Rosenblum}, \citenamefont {Peer},\
  and\ \citenamefont {L{\'e}vy}}]{Bogart:2014}%
  \BibitemOpen
  \bibfield  {author} {\bibinfo {author} {\bibfnamefont {L.~K.}\ \bibnamefont
  {Bogart}}, \bibinfo {author} {\bibfnamefont {G.}~\bibnamefont {Pourroy}},
  \bibinfo {author} {\bibfnamefont {C.~J.}\ \bibnamefont {Murphy}}, \bibinfo
  {author} {\bibfnamefont {V.}~\bibnamefont {Puntes}}, \bibinfo {author}
  {\bibfnamefont {T.}~\bibnamefont {Pellegrino}}, \bibinfo {author}
  {\bibfnamefont {D.}~\bibnamefont {Rosenblum}}, \bibinfo {author}
  {\bibfnamefont {D.}~\bibnamefont {Peer}}, \ and\ \bibinfo {author}
  {\bibfnamefont {R.}~\bibnamefont {L{\'e}vy}},\ }\href@noop {} {\bibfield
  {journal} {\bibinfo  {journal} {ACS Nano}\ }\textbf {\bibinfo {volume} {8}},\
  \bibinfo {pages} {3107} (\bibinfo {year} {2014})}\BibitemShut {NoStop}%
\bibitem [{\citenamefont {Whitesides}\ and\ \citenamefont
  {Grzybowski}(2002)}]{Whitesides:2002}%
  \BibitemOpen
  \bibfield  {author} {\bibinfo {author} {\bibfnamefont {G.~M.}\ \bibnamefont
  {Whitesides}}\ and\ \bibinfo {author} {\bibfnamefont {B.}~\bibnamefont
  {Grzybowski}},\ }\href@noop {} {\bibfield  {journal} {\bibinfo  {journal}
  {Science}\ }\textbf {\bibinfo {volume} {295}},\ \bibinfo {pages} {2418}
  (\bibinfo {year} {2002})}\BibitemShut {NoStop}%
\bibitem [{\citenamefont {Moran}\ and\ \citenamefont
  {Posner}(2017)}]{Moran:2017}%
  \BibitemOpen
  \bibfield  {author} {\bibinfo {author} {\bibfnamefont {J.~L.}\ \bibnamefont
  {Moran}}\ and\ \bibinfo {author} {\bibfnamefont {J.~D.}\ \bibnamefont
  {Posner}},\ }\href@noop {} {\bibfield  {journal} {\bibinfo  {journal} {Annu.
  Rev. Fluid Mech.}\ }\textbf {\bibinfo {volume} {49}},\ \bibinfo {pages} {511}
  (\bibinfo {year} {2017})}\BibitemShut {NoStop}%
\bibitem [{\citenamefont {Kroy}\ \emph {et~al.}(2016)\citenamefont {Kroy},
  \citenamefont {Chakraborty},\ and\ \citenamefont {Cichos}}]{Kroy:2016}%
  \BibitemOpen
  \bibfield  {author} {\bibinfo {author} {\bibfnamefont {K.}~\bibnamefont
  {Kroy}}, \bibinfo {author} {\bibfnamefont {D.}~\bibnamefont {Chakraborty}}, \
  and\ \bibinfo {author} {\bibfnamefont {F.}~\bibnamefont {Cichos}},\
  }\href@noop {} {\bibfield  {journal} {\bibinfo  {journal} {Eur. Phys. J.
  Spec. Top.}\ }\textbf {\bibinfo {volume} {225}},\ \bibinfo {pages} {2207}
  (\bibinfo {year} {2016})}\BibitemShut {NoStop}%
\bibitem [{\citenamefont {Marchetti}\ \emph {et~al.}(2013)\citenamefont
  {Marchetti}, \citenamefont {Joanny}, \citenamefont {Ramaswamy}, \citenamefont
  {Liverpool}, \citenamefont {Prost}, \citenamefont {Rao},\ and\ \citenamefont
  {Simha}}]{Marchetti:2013}%
  \BibitemOpen
  \bibfield  {author} {\bibinfo {author} {\bibfnamefont {M.~C.}\ \bibnamefont
  {Marchetti}}, \bibinfo {author} {\bibfnamefont {J.~F.}\ \bibnamefont
  {Joanny}}, \bibinfo {author} {\bibfnamefont {S.}~\bibnamefont {Ramaswamy}},
  \bibinfo {author} {\bibfnamefont {T.~B.}\ \bibnamefont {Liverpool}}, \bibinfo
  {author} {\bibfnamefont {J.}~\bibnamefont {Prost}}, \bibinfo {author}
  {\bibfnamefont {M.}~\bibnamefont {Rao}}, \ and\ \bibinfo {author}
  {\bibfnamefont {R.~A.}\ \bibnamefont {Simha}},\ }\href@noop {} {\bibfield
  {journal} {\bibinfo  {journal} {Rev. Mod. Phys.}\ }\textbf {\bibinfo {volume}
  {85}},\ \bibinfo {pages} {1143} (\bibinfo {year} {2013})}\BibitemShut
  {NoStop}%
\bibitem [{\citenamefont {Singh}\ and\ \citenamefont
  {Adhikari}(2016)}]{Singh:2016}%
  \BibitemOpen
  \bibfield  {author} {\bibinfo {author} {\bibfnamefont {R.}~\bibnamefont
  {Singh}}\ and\ \bibinfo {author} {\bibfnamefont {R.}~\bibnamefont
  {Adhikari}},\ }\href@noop {} {\bibfield  {journal} {\bibinfo  {journal}
  {Phys. Rev. Lett.}\ }\textbf {\bibinfo {volume} {117}},\ \bibinfo {pages}
  {228002} (\bibinfo {year} {2016})}\BibitemShut {NoStop}%
\bibitem [{\citenamefont {Shen}\ \emph {et~al.}(2019)\citenamefont {Shen},
  \citenamefont {Würger},\ and\ \citenamefont {Lintuvuori}}]{Shen:2019}%
  \BibitemOpen
  \bibfield  {author} {\bibinfo {author} {\bibfnamefont {Z.}~\bibnamefont
  {Shen}}, \bibinfo {author} {\bibfnamefont {A.}~\bibnamefont {Würger}}, \
  and\ \bibinfo {author} {\bibfnamefont {J.~S.}\ \bibnamefont {Lintuvuori}},\
  }\href@noop {} {\bibfield  {journal} {\bibinfo  {journal} {Soft Matter}\
  }\textbf {\bibinfo {volume} {15}},\ \bibinfo {pages} {1508} (\bibinfo {year}
  {2019})}\BibitemShut {NoStop}%
\bibitem [{\citenamefont {R\"uckner}\ and\ \citenamefont
  {Kapral}(2007)}]{Gunnar:2007}%
  \BibitemOpen
  \bibfield  {author} {\bibinfo {author} {\bibfnamefont {G.}~\bibnamefont
  {R\"uckner}}\ and\ \bibinfo {author} {\bibfnamefont {R.}~\bibnamefont
  {Kapral}},\ }\href@noop {} {\bibfield  {journal} {\bibinfo  {journal} {Phys.
  Rev. Lett.}\ }\textbf {\bibinfo {volume} {98}},\ \bibinfo {pages} {150603}
  (\bibinfo {year} {2007})}\BibitemShut {NoStop}%
\bibitem [{\citenamefont {Valadares}\ \emph {et~al.}(2010)\citenamefont
  {Valadares}, \citenamefont {Tao}, \citenamefont {Zacharia}, \citenamefont
  {Kitaev}, \citenamefont {Galembeck}, \citenamefont {Kapral},\ and\
  \citenamefont {Ozin}}]{Valadares:2010}%
  \BibitemOpen
  \bibfield  {author} {\bibinfo {author} {\bibfnamefont {L.~F.}\ \bibnamefont
  {Valadares}}, \bibinfo {author} {\bibfnamefont {Y.-G.}\ \bibnamefont {Tao}},
  \bibinfo {author} {\bibfnamefont {N.~S.}\ \bibnamefont {Zacharia}}, \bibinfo
  {author} {\bibfnamefont {V.}~\bibnamefont {Kitaev}}, \bibinfo {author}
  {\bibfnamefont {F.}~\bibnamefont {Galembeck}}, \bibinfo {author}
  {\bibfnamefont {R.}~\bibnamefont {Kapral}}, \ and\ \bibinfo {author}
  {\bibfnamefont {G.~A.}\ \bibnamefont {Ozin}},\ }\href@noop {} {\bibfield
  {journal} {\bibinfo  {journal} {Small}\ }\textbf {\bibinfo {volume} {6}},\
  \bibinfo {pages} {565} (\bibinfo {year} {2010})}\BibitemShut {NoStop}%
\bibitem [{\citenamefont {Anderson}(1989)}]{Anderson:1989}%
  \BibitemOpen
  \bibfield  {author} {\bibinfo {author} {\bibfnamefont {J.~L.}\ \bibnamefont
  {Anderson}},\ }\href@noop {} {\bibfield  {journal} {\bibinfo  {journal} {Ann.
  Rev. Fluid. Mech.}\ }\textbf {\bibinfo {volume} {21}},\ \bibinfo {pages} {61}
  (\bibinfo {year} {1989})}\BibitemShut {NoStop}%
\bibitem [{\citenamefont {Kapral}(2008)}]{Raymond:2008}%
  \BibitemOpen
  \bibfield  {author} {\bibinfo {author} {\bibfnamefont {R.}~\bibnamefont
  {Kapral}},\ }\href@noop {} {\bibfield  {journal} {\bibinfo  {journal} {Adv.
  Chem. Phys.}\ }\textbf {\bibinfo {volume} {140}},\ \bibinfo {pages} {89}
  (\bibinfo {year} {2008})}\BibitemShut {NoStop}%
\bibitem [{\citenamefont {Gompper}\ \emph {et~al.}(2009)\citenamefont
  {Gompper}, \citenamefont {Ihle}, \citenamefont {Kroll},\ and\ \citenamefont
  {Winkler}}]{Gompper:2009}%
  \BibitemOpen
  \bibfield  {author} {\bibinfo {author} {\bibfnamefont {G.}~\bibnamefont
  {Gompper}}, \bibinfo {author} {\bibfnamefont {T.}~\bibnamefont {Ihle}},
  \bibinfo {author} {\bibfnamefont {D.~M.}\ \bibnamefont {Kroll}}, \ and\
  \bibinfo {author} {\bibfnamefont {R.~G.}\ \bibnamefont {Winkler}},\ }\enquote
  {\bibinfo {title} {Multi-particle collision dynamics: A particle-based
  mesoscale simulation approach to the hydrodynamics of complex fluids},}\ \
  (\bibinfo  {publisher} {Springer Berlin Heidelberg},\ \bibinfo {year}
  {2009})\ pp.\ \bibinfo {pages} {1--87}\BibitemShut {NoStop}%
\bibitem [{\citenamefont {Ihle}\ and\ \citenamefont {Kroll}(2001)}]{Ihle:2001}%
  \BibitemOpen
  \bibfield  {author} {\bibinfo {author} {\bibfnamefont {T.}~\bibnamefont
  {Ihle}}\ and\ \bibinfo {author} {\bibfnamefont {D.~M.}\ \bibnamefont
  {Kroll}},\ }\href@noop {} {\bibfield  {journal} {\bibinfo  {journal} {Phys.
  Rev. E}\ }\textbf {\bibinfo {volume} {63}},\ \bibinfo {pages} {020201}
  (\bibinfo {year} {2001})}\BibitemShut {NoStop}%
\bibitem [{\citenamefont {Golestanian}\ \emph {et~al.}(2007)\citenamefont
  {Golestanian}, \citenamefont {Liverpool},\ and\ \citenamefont
  {Ajdari}}]{Golestanian:2007}%
  \BibitemOpen
  \bibfield  {author} {\bibinfo {author} {\bibfnamefont {R.}~\bibnamefont
  {Golestanian}}, \bibinfo {author} {\bibfnamefont {T.~B.}\ \bibnamefont
  {Liverpool}}, \ and\ \bibinfo {author} {\bibfnamefont {A.}~\bibnamefont
  {Ajdari}},\ }\href@noop {} {\bibfield  {journal} {\bibinfo  {journal} {New J.
  Phys.}\ }\textbf {\bibinfo {volume} {9}},\ \bibinfo {pages} {126} (\bibinfo
  {year} {2007})}\BibitemShut {NoStop}%
\bibitem [{\citenamefont {Reigh}\ and\ \citenamefont
  {Kapral}(2015)}]{Reigh:2015}%
  \BibitemOpen
  \bibfield  {author} {\bibinfo {author} {\bibfnamefont {S.~Y.}\ \bibnamefont
  {Reigh}}\ and\ \bibinfo {author} {\bibfnamefont {R.}~\bibnamefont {Kapral}},\
  }\href@noop {} {\bibfield  {journal} {\bibinfo  {journal} {Soft Matter}\
  }\textbf {\bibinfo {volume} {11}} (\bibinfo {year} {2015})}\BibitemShut
  {NoStop}%
\bibitem [{\citenamefont {Singh}\ \emph {et~al.}(2017)\citenamefont {Singh},
  \citenamefont {Choudhury}, \citenamefont {Fischer},\ and\ \citenamefont
  {Mark}}]{Singh:2017}%
  \BibitemOpen
  \bibfield  {author} {\bibinfo {author} {\bibfnamefont {D.~P.}\ \bibnamefont
  {Singh}}, \bibinfo {author} {\bibfnamefont {U.}~\bibnamefont {Choudhury}},
  \bibinfo {author} {\bibfnamefont {P.}~\bibnamefont {Fischer}}, \ and\
  \bibinfo {author} {\bibfnamefont {A.~G.}\ \bibnamefont {Mark}},\ }\href@noop
  {} {\bibfield  {journal} {\bibinfo  {journal} {Adv. Mater.}\ }\textbf
  {\bibinfo {volume} {29}},\ \bibinfo {pages} {1701328} (\bibinfo {year}
  {2017})}\BibitemShut {NoStop}%
\bibitem [{\citenamefont {Gonzalez}\ and\ \citenamefont
  {Soto}(2019)}]{Gonzalez:2019}%
  \BibitemOpen
  \bibfield  {author} {\bibinfo {author} {\bibfnamefont {S.}~\bibnamefont
  {Gonzalez}}\ and\ \bibinfo {author} {\bibfnamefont {R.}~\bibnamefont
  {Soto}},\ }\href@noop {} {\bibfield  {journal} {\bibinfo  {journal} {New J.
  Phys.}\ }\textbf {\bibinfo {volume} {21}},\ \bibinfo {pages} {033041}
  (\bibinfo {year} {2019})}\BibitemShut {NoStop}%
\bibitem [{\citenamefont {Peruani}\ \emph {et~al.}(2006)\citenamefont
  {Peruani}, \citenamefont {Deutsch},\ and\ \citenamefont {B\"ar}}]{Bar:2006}%
  \BibitemOpen
  \bibfield  {author} {\bibinfo {author} {\bibfnamefont {F.}~\bibnamefont
  {Peruani}}, \bibinfo {author} {\bibfnamefont {A.}~\bibnamefont {Deutsch}}, \
  and\ \bibinfo {author} {\bibfnamefont {M.}~\bibnamefont {B\"ar}},\
  }\href@noop {} {\bibfield  {journal} {\bibinfo  {journal} {Phys. Rev. E}\
  }\textbf {\bibinfo {volume} {74}},\ \bibinfo {pages} {030904} (\bibinfo
  {year} {2006})}\BibitemShut {NoStop}%
\bibitem [{\citenamefont {Sawin}\ \emph {et~al.}(1992)\citenamefont {Sawin},
  \citenamefont {LeGuellec}, \citenamefont {Philippe},\ and\ \citenamefont
  {Mitchison}}]{Sawin:1992}%
  \BibitemOpen
  \bibfield  {author} {\bibinfo {author} {\bibfnamefont {K.~E.}\ \bibnamefont
  {Sawin}}, \bibinfo {author} {\bibfnamefont {K.}~\bibnamefont {LeGuellec}},
  \bibinfo {author} {\bibfnamefont {M.}~\bibnamefont {Philippe}}, \ and\
  \bibinfo {author} {\bibfnamefont {T.~J.}\ \bibnamefont {Mitchison}},\
  }\href@noop {} {\bibfield  {journal} {\bibinfo  {journal} {Nature}\ }\textbf
  {\bibinfo {volume} {359}},\ \bibinfo {pages} {540} (\bibinfo {year}
  {1992})}\BibitemShut {NoStop}%
\end{thebibliography}%
\end{document}